\PassOptionsToPackage{unicode}{hyperref}
\PassOptionsToPackage{hyphens}{url}
\PassOptionsToPackage{dvipsnames,svgnames,x11names}{xcolor}
\documentclass[
  12pt]{article}

\usepackage{amsmath,amssymb}
\usepackage{iftex}
\ifPDFTeX
  \usepackage[T1]{fontenc}
  \usepackage[utf8]{inputenc}
  \usepackage{textcomp} % provide euro and other symbols
\else % if luatex or xetex
  \usepackage{unicode-math}
  \defaultfontfeatures{Scale=MatchLowercase}
  \defaultfontfeatures[\rmfamily]{Ligatures=TeX,Scale=1}
\fi
\usepackage{lmodern}
\ifPDFTeX\else  
\fi
\IfFileExists{upquote.sty}{\usepackage{upquote}}{}
\IfFileExists{microtype.sty}{% use microtype if available
  \usepackage[]{microtype}
  \UseMicrotypeSet[protrusion]{basicmath} % disable protrusion for tt fonts
}{}
\makeatletter
\@ifundefined{KOMAClassName}{% if non-KOMA class
  \IfFileExists{parskip.sty}{%
    \usepackage{parskip}
  }{% else
    \setlength{\parindent}{0pt}
    \setlength{\parskip}{6pt plus 2pt minus 1pt}}
}{% if KOMA class
  \KOMAoptions{parskip=half}}
\makeatother
\usepackage{xcolor}
\makeatletter
\ifx\paragraph\undefined\else
  \let\oldparagraph\paragraph
  \renewcommand{\paragraph}{
    \@ifstar
      \xxxParagraphStar
      \xxxParagraphNoStar
  }
  \newcommand{\xxxParagraphStar}[1]{\oldparagraph*{#1}\mbox{}}
  \newcommand{\xxxParagraphNoStar}[1]{\oldparagraph{#1}\mbox{}}
\fi
\ifx\subparagraph\undefined\else
  \let\oldsubparagraph\subparagraph
  \renewcommand{\subparagraph}{
    \@ifstar
      \xxxSubParagraphStar
      \xxxSubParagraphNoStar
  }
  \newcommand{\xxxSubParagraphStar}[1]{\oldsubparagraph*{#1}\mbox{}}
  \newcommand{\xxxSubParagraphNoStar}[1]{\oldsubparagraph{#1}\mbox{}}
\fi
\makeatother

\usepackage{longtable,booktabs,array}
\usepackage{calc} % for calculating minipage widths
\usepackage{etoolbox}
\makeatletter
\patchcmd\longtable{\par}{\if@noskipsec\mbox{}\fi\par}{}{}
\makeatother
\IfFileExists{footnotehyper.sty}{\usepackage{footnotehyper}}{\usepackage{footnote}}
\makesavenoteenv{longtable}
\usepackage{graphicx}
\makeatletter
\def\maxwidth{\ifdim\Gin@nat@width>\linewidth\linewidth\else\Gin@nat@width\fi}
\def\maxheight{\ifdim\Gin@nat@height>\textheight\textheight\else\Gin@nat@height\fi}
\makeatother
\setkeys{Gin}{width=\maxwidth,height=\maxheight,keepaspectratio}
\makeatletter
\def\fps@figure{htbp}
\makeatother

\makeatletter
\@ifpackageloaded{caption}{}{\usepackage{caption}}
\AtBeginDocument{%
\ifdefined\contentsname
  \renewcommand*\contentsname{Table of contents}
\else
  \newcommand\contentsname{Table of contents}
\fi
\ifdefined\listfigurename
  \renewcommand*\listfigurename{List of Figures}
\else
  \newcommand\listfigurename{List of Figures}
\fi
\ifdefined\listtablename
  \renewcommand*\listtablename{List of Tables}
\else
  \newcommand\listtablename{List of Tables}
\fi
\ifdefined\figurename
  \renewcommand*\figurename{Figure}
\else
  \newcommand\figurename{Figure}
\fi
\ifdefined\tablename
  \renewcommand*\tablename{Table}
\else
  \newcommand\tablename{Table}
\fi
}
\@ifpackageloaded{float}{}{\usepackage{float}}
\floatstyle{ruled}
\@ifundefined{c@chapter}{\newfloat{codelisting}{h}{lop}}{\newfloat{codelisting}{h}{lop}[chapter]}
\floatname{codelisting}{Listing}

\makeatother
\makeatletter
\@ifpackageloaded{caption}{}{\usepackage{caption}}
\@ifpackageloaded{subcaption}{}{\usepackage{subcaption}}
\makeatother

\ifLuaTeX
  \usepackage{selnolig}  % disable illegal ligatures
\fi
\usepackage[]{natbib}
\usepackage{bookmark}

\IfFileExists{xurl.sty}{\usepackage{xurl}}{} % add URL line breaks if available
\hypersetup{
  pdftitle={Title},
  pdfauthor={Author 1; Author 2},
  pdfkeywords={3 to 6 keywords, that do not appear in the title},
  colorlinks=true,
  linkcolor={blue},
  filecolor={Maroon},
  citecolor={Blue},
  urlcolor={Blue},
  pdfcreator={LaTeX via pandoc}}

\newcommand{\anon}{1}

\usepackage{lipsum}
\usepackage{mathtools, amssymb, amsmath,amsthm }
\usepackage{float} 
\usepackage{hyperref,bookmark}
\usepackage{enumerate}
\usepackage{xcolor}
\usepackage{setspace}
\allowdisplaybreaks

\usepackage{xpatch}
 
\newcommand{\indep}{\perp \!\!\! \perp}
\newtheorem{thm}{Theorem}
\newtheorem*{thm*}{Theorem}
\newtheorem{cor}{Corollary}
\newtheorem{lemma}{Lemma}

\begin{document}

\def\spacingset#1{\renewcommand{\baselinestretch}%
{#1}\small\normalsize} \spacingset{1}

%%%%%%%%%%%%%%%%%%%%%%%%%%%%%%%%%%%%%%%%%%%%%%%%%%%%%%%%%%%%%%%%%%%%%%%%%%%%%%

\if1\anon
{
  \title{Subspace Ordering for Maximum Response Preservation in Sufficient Dimension Reduction}
  \author{Derik T. Boonstra, Rakheon Kim, and Dean M. Young\\
    Department of Statistical Science, Baylor University\\
    }
    \date{}
  \maketitle
} \fi

\if0\anon
{
  \bigskip
  \bigskip
  \bigskip
  \begin{center}
    {\LARGE Subspace Ordering for Maximum Response Preservation in Sufficient Dimension Reduction}
\end{center}
  \medskip
} \fi

\bigskip
\begin{abstract}
% The text of your abstract. 200 or fewer words.
Sufficient dimension reduction (SDR) methods aim to identify a dimension reduction subspace (DRS) that preserves all the information about the conditional distribution of a response given its predictor. Traditional SDR methods determine the DRS by solving a method-specific generalized eigenvalue problem and selecting the eigenvectors corresponding to the largest eigenvalues. In this article, we argue against the long-standing convention of using eigenvalues as the measure of subspace importance and propose alternative ordering criteria that directly assess the predictive relevance of each subspace. For a binary response, we introduce a subspace ordering criterion based on the absolute value of the independent Student’s T-statistic. Theoretically, our criterion identifies subspaces that achieve the local minimum Bayes' error rate and yields consistent ordering of directions under mild regularity conditions. Additionally, we  employ an F-statistic to provide a framework that unifies categorical and continuous responses under a single subspace criterion. We evaluate our proposed criteria within multiple SDR methods through extensive simulation studies and applications to real-data. Our empirical results demonstrate the efficacy of reordering subspaces using our proposed criteria, which generally yields significant improvements in classification accuracy and subspace estimation compared to ordering by eigenvalues. 
\end{abstract}

\noindent%

\spacingset{1.8} % DON'T change the spacing!

\section{Introduction}\label{sec:1_intro}

The \textit{dimension reduction subspace} (\textit{DRS}) in \textit{sufficient dimension reduction} (\textit{SDR}) is often composed of a subset of the leading eigenvectors of a method-specific generalized eigenvalue problem. That is, the \textit{DRS} is spanned by the eigenvectors corresponding to the largest eigenvalues in magnitude. This approach to choosing eigenvectors based upon eigenvalues is common practice and has the intent of maximizing the variability of the data in the chosen subspace. However, for supervised statistical learning, we argue that the use of eigenvalues to determine a relevant \textit{DRS} is generally flawed. This problem results because maximum variability does not guarantee that the selected subspace best preserves the relationship between the predictors and the response, which is the primary goal of \textit{SDR}. 

Let \(Y\) be the response of the predictor \(\mathbf{X} \in \mathbb{R}^{p}\). Then, the goal of \textit{SDR} is to project \(\mathbf{X}\) onto the smallest possible subspace \(\mathcal{S} \subseteq \mathbb{R}^{p}\) without any loss of information with respect to \(Y|\mathbf{X}\). In \textit{SDR}, the dimension reduction is usually constrained to a linear transformation \(\mathbf{P}_{\mathcal{S}}\mathbf{X}\), where \(\mathbf{P}_{\mathcal{S}} \in \mathbb{R}^{p \times p}\) is the projection matrix onto \(\mathcal{S}\) in the standard inner product. Let ``\(\sim\)'' and ``\(\indep\)'' denote ``distributed as'' and ``independent of,'' respectively. We formally define the dimension reduction as a \textit{sufficient linear reduction} if it satisfies at least one of the following: (i) \(Y \indep \mathbf{X}|\mathbf{P}_{\mathcal{S}} \mathbf{X}\), (ii) \(\mathbf{X}|\left(Y,\mathbf{P}_{\mathcal{S}}\mathbf{X}\right) \sim \mathbf{X}|\mathbf{P}_{\mathcal{S}}\mathbf{X}\), or (iii) \(Y|\mathbf{X} \sim Y|\mathbf{P}_{\mathcal{S}}\mathbf{X}\). Thus, \(\mathcal{S}\) is called a \textit{DRS} and is said to be a \textit{minimum DRS} for \(Y|\mathbf{X}\) if \(\text{dim}(\mathcal{S}) \leq \text{dim}(\mathcal{S}_{DRS})\), where \(\text{dim}(\cdot)\) denotes dimension and \(\mathcal{S}_{DRS}\) represents all other \textit{DRS}. The \textit{minimum DRS}, however, may not be unique (e.g., see Section 6.3 of \cite{cook1998}). To address this issue, \cite{cook1998} introduced the \textit{central subspace} (\textit{CS}), defined as the intersection of all \textit{DRS}, and denoted it as \(\mathcal{S}_{Y|\mathbf{X}} = \cap \mathcal{S}_{DRS}\). Given that \(\cap \mathcal{S}_{DRS}\) itself is a \textit{DRS} and, under mild conditions given by \cite{cook1998}, \(\mathcal{S}_{Y|\mathbf{X}}\) exists and is the \textit{unique minimum DRS}. Thus, most \textit{SDR} methods estimate the \textit{CS}, or at least a portion of the \textit{CS}, as the target subspace. Throughout the paper, we assume the existence of the \textit{CS}. Let \(d < p\) be the structural dimension of the \textit{CS}, and let \(\boldsymbol{\beta} = \left(\boldsymbol{\beta}_{1}, \ldots, \boldsymbol{\beta}_{d}\right) \in \mathbb{R}^{p \times d}\) be a basis matrix of \(\mathcal{S}_{Y|\mathbf{X}}\). Thus, \textit{SDR} methods can reduce the dimensionality of predictors to \(\boldsymbol{\beta}^{\top}\mathbf{X} \in \mathbb{R}^{d}\) for subsequent supervised learning without loss of information. For a comprehensive review of \textit{SDR}, see \cite{li2018}.

Most classical \textit{SDR} methods rely on slicing the data into \(H\) contiguous non-overlapping intervals to construct functions of the first two conditional moments with the goal of recovering the \textit{CS}. Notable examples of this approach include \textit{sliced inverse regression} (\textit{SIR}), introduced by \cite{li1991}, and \textit{sliced average variance estimation} (\textit{SAVE}), proposed by \cite{cook1991}. This slicing-based framework is particularly suited to a continuous response, where slicing facilitates tractable estimation of conditional expectations and yields a \textit{DRS} that is contained in \(\mathcal{S}_{Y|\mathbf{X}}\) for sufficiently large \(H\). When the response is categorical, slicing becomes trivial since the grouping of observations is explicit by the \(H\) distinct populations. Moreover, with a categorical response, the emphasis is often on optimizing a classification rule rather than examining other aspects of \(Y|\mathbf{X}\). Thus, \cite{cook2001} proposed the \textit{central discriminant subspace} (\textit{CDS}) as an alternative target subspace to the \textit{CS} for discriminant analysis. Let the Bayes' rule be \(\phi(\mathbf{X}) \coloneqq \arg \max _{h = 1, \ldots, H} \Pr\left(Y =  h | \mathbf{X}\right)\). For a subspace \(\mathcal{S} \subseteq \mathbb{R}^{p}\), let  \(\phi_{\mathcal{S}}(\mathbf{X}) \coloneqq \arg \max _{h = 1, \ldots, H} \Pr\left(Y =  h |\mathbf{P}_{\mathcal{S}} \mathbf{X}\right)\). Moreover, for any basis matrix \(\boldsymbol{\beta}\) such that \(\text{span}(\boldsymbol{\beta}) = \mathcal{S}\), we have \(\phi_{\mathcal{S}}(\mathbf{X}) = \arg \max _{h = 1, \ldots, H} \Pr\left(Y =  h |\boldsymbol{\beta}^{\top}\mathbf{X}\right)\). Thus, if \(\mathcal{S}\) satisfies \(\phi_{\mathcal{S}}(\mathbf{X}) = \phi(\mathbf{X})\), then it is a \textit{discriminant subspace}. The \textit{CDS}, denoted as \(\mathcal{S}_{D(Y|\mathbf{X})} \subseteq \mathcal{S}_{Y|\mathbf{X}}\), is then defined as the intersection of all \textit{discriminant subspaces}, given that the intersection itself is a \textit{discriminant subspace}.

Regardless of whether the emphasis is on the \textit{CS} or \textit{CDS}, most \textit{SDR} methods rely on eigenvectors to estimate \(\text{span}(\boldsymbol{\beta})\). These eigenvectors are typically ordered by the magnitude of their associated eigenvalues with the assumption that subspaces corresponding to relatively large eigenvalues are more informative. However, large eigenvalues generally only represent larger data variability in the subspace and do not guarantee that the predictor-response relationship is preserved (e.g., see \cite{huber1985}). Thus, we propose new and more appropriate subspace ordering criteria that explicitly capture the predictive information in each direction, thereby ensuring that the leading subspaces align with the intended goal of the supervised learner.

More specifically, when the response is binary, we propose using the absolute value of an independent \textit{Student’s T-statistic}, introduced by \cite{welch1947}, as a simple yet effective importance measure for a \textit{DRS}. We naturally extend the idea of a subspace ordering criterion to a categorical response with more than two populations by employing an \textit{F-statistic}, introduced by \cite{fisher1925}. Furthermore, within the slicing-based framework, we demonstrate that an \textit{F-statistic} can also serve as a subspace criterion in the continuous response setting. That is, by emphasizing maximum slice separation, we can often identify subspaces that best capture the conditional moments. Although methods exist to determine the structural dimension \(d\) of the \textit{CS} (e.g., see Chapters 9 and 10 in \cite{li2018} and see \cite{zeng2024}), to the best of our knowledge, we are the first to propose reordering the \textit{DRS} by a criterion different from the eigenvalue magnitudes. We believe that, while the intuition of these proposed criteria is simple, its simplicity offers a novel and interpretable importance measure for subspaces that addresses a previously overlooked but crucial aspect of subspace selection---namely, the need to directly assess predictive relevance rather than defaulting to eigenvalue magnitude.

The remainder of the paper proceeds as follows. In Section \ref{sec:2_review}, we establish the notation used throughout the paper and provide a brief review of relevant methodologies. In Section \ref{sec:3_T}, we provide a simple example to illustrate the potential gain of reordering a \textit{DRS} by an \textit{Student's T-statistic} rather than the eigenvalue magnitude. We then study the theoretical properties of the criterion to establish it as a consistent and relevant importance measure for a \textit{DRS}. In Section \ref{sec:4_F}, we establish a criterion using an \textit{F-statistic} for a categorical or continuous response. In Sections \ref{sec:5_simulation} and \ref{sec:6_real_data_apps}, we present the simulation studies and real-data applications, respectively. In Section \ref{sec:discussion}, we discuss additional work and conclude our findings.
 
\section{A Review of Methodologies}\label{sec:2_review}

\subsection{Notation}
First, we define the following notation that will be used throughout the paper. Let \(\mathbb{R}^{m \times n}\) represent the set of all \(m \times n\) real matrices, \(\mathbb{S}^{p} \subset \mathbb{R}^{p \times p}\) represent the set of \(p \times p\) real symmetric matrices, and \(\mathbb{S}^{p}_{+} \subset \mathbb{S}^p\) denote the interior of the cone of \(p \times p\) real symmetric positive-definite matrices. Let \(\oplus\) denote the direct sum such that \(\mathbf{A} \oplus \mathbf{B} = \begin{bsmallmatrix} \mathbf{A} & \mathbf{0} \\ \mathbf{0} & \mathbf{B}\end{bsmallmatrix}\). 
% Let \(\mathbf{X}\) be a random vector defined on the probability space \((\Omega, \mathcal{F}, P)\). 
For \(h = 1, \ldots, H\), let \(\Pi_h\) represent the \(h^{\text{th}}\) distinct interval of \(Y\) with the \textit{a priori} membership \(\pi_h = \Pr(Y = h)\), where \(\sum_{h = 1}^{H}\pi_{h} = 1\). Let \(\boldsymbol{\Sigma}_{h} \in \mathbb{S}^{p}_{+}\) denote the population covariance matrix for \(\Pi_h\), and let \(\boldsymbol{\mu}_{h} \in \mathbb{R}^{p\times 1}\) be the population mean vector for \(\Pi_h\).

\subsection{Discriminant Analysis}

In Section \ref{sec:1_intro}, we defined the \textit{CDS} as the intersection of all subspaces that preserves Bayes' rule, \( \phi(\mathbf{X}) = \arg\max_{h=1,\ldots,H} \Pr(Y = h \mid \mathbf{X}) \). Because the conditional class probability \( \Pr(Y = h \mid \mathbf{X}) \) is generally unknown and lacks a closed-form expression, a common approach is to consider settings in which it is tractable. One such setting is the multivariate normal population model for the predictor \( \mathbf{X} \in \mathbb{R}^p \) given the categorical response \( Y \in \{1, \ldots, H\} \), \( H \geq 2 \), which is given by 
\begin{equation}\label{eq:model}
    \mathbf{X} \mid \{Y = h\} \sim \mathcal{N}(\boldsymbol{\mu}_h, \boldsymbol{\Sigma}_h), \quad h = 1, \ldots, H.
\end{equation}
If no assumption is made on \(\boldsymbol{\Sigma}_{1}, \ldots, \boldsymbol{\Sigma}_{H}\), then, under model \eqref{eq:model}, the Bayes' rule for classification is
\begin{equation}\label{eq:qda_og}
\phi_{QDA}(\mathbf{X}) = \arg\max_{h=1,\ldots,H} \left\{ \log \pi_h + \frac{1}{2} \log |\boldsymbol{\Sigma}_h| + \frac{1}{2} (\mathbf{X} - \boldsymbol{\mu}_h)^{\top} \boldsymbol{\Sigma}_h^{-1} (\mathbf{X} - \boldsymbol{\mu}_h) \right\},
\end{equation}
which is commonly referred to as the Bayes' \textit{quadratic discriminant analysis} (\textit{QDA}) rule.

Because the \textit{QDA} rule does not require homoscedasticity, the Bayes' rule for classification is a quadratic function of \(\mathbf{X}\). Although expression \eqref{eq:qda_og} gives the Bayes' rule in its standard form, we can algebraically reformulate it to explicitly isolate
the linear and quadratic components by subtracting the discriminant function for a reference class (e.g., class 1),
\begin{equation}\label{eq:qda}
\phi_{QDA}(\mathbf{X}) = \arg\max_{h=1,\ldots,H} \left\{ c_h - \mathbf{X}^{\top}(\boldsymbol{\Sigma}_h^{-1} \boldsymbol{\mu}_h - \boldsymbol{\Sigma}_1^{-1} \boldsymbol{\mu}_1) + \frac{1}{2} \mathbf{X}^{\top}(\boldsymbol{\Sigma}_h^{-1} - \boldsymbol{\Sigma}_1^{-1}) \mathbf{X} \right\},
\end{equation}
where \( c_h = \log \pi_h + \log|\boldsymbol{\Sigma}_h|/2 + \boldsymbol{\mu}_h^{\top}\boldsymbol{\Sigma}_h^{-1}\boldsymbol{\mu}_h/2 \) is a constant term that does not depend on \(\mathbf{X}\). Thus, expression \eqref{eq:qda} yields that optimal dimension reduction under the \textit{QDA} model is characterized by the subspace 
\begin{equation}\label{eq:qda_L}
\mathcal{L} \coloneqq \text{span}\{\boldsymbol{\Sigma}_h^{-1} \boldsymbol{\mu}_h - \boldsymbol{\Sigma}_1^{-1} \boldsymbol{\mu}_1 \mid h = 2, \ldots, H\} \subseteq \mathbb{R}^p,
\end{equation}
which corresponds to the linear components, and 
\begin{equation}\label{eq:qda_Q}
    \mathcal{Q} \coloneqq \text{span}\{\boldsymbol{\Sigma}_h^{-1} - \boldsymbol{\Sigma}_1^{-1} \mid h = 2, \ldots, H\} \subseteq \mathbb{R}^{p},
\end{equation}
which corresponds to the quadratic components. This result is formalized in the following lemma. All proofs for lemmas and theorems are given in the Supplementary Material.

\vspace{1em}
\begin{lemma}\label{lem:qda_cds}
Let \(\mathcal{L}\) and \(\mathcal{Q}\) be as defined in \eqref{eq:qda_L} and \eqref{eq:qda_Q}, respectively. Then, under model \eqref{eq:model}, \(\mathcal{S}_{Y \mid \mathbf{X}} = \mathcal{S}_{D(Y \mid \mathbf{X})} = \mathcal{L} \cup \mathcal{Q}\). 
\end{lemma}
\vspace{1em}

If we assume \(\boldsymbol{\Sigma}_{1} = \ldots = \boldsymbol{\Sigma}_{H} = \boldsymbol{\Sigma}\), then the Bayes' rule simplifies to what is commonly referred to as the Bayes' \textit{linear discriminant analysis} (\textit{LDA}) rule, which reduces \eqref{eq:qda} to
\begin{equation}\label{eq:lda}
    \phi_{LDA}(\mathbf{X}) = \arg\max_{h = 1, \ldots, H} \left\{ \log (\pi_{h}/\pi_{1}) +  (\boldsymbol{\mu}_{h} -  \boldsymbol{\mu}_{1})^{\top}\boldsymbol{\Sigma}^{-1}(\mathbf{X} - [\boldsymbol{\mu}_{h} +  \boldsymbol{\mu}_{1}]/2) \right\}.
\end{equation}
To distinguish the \(H\) populations under such a linear model, we require at most \(H - 1\) directions. Thus, we consider the subspace 
\begin{equation}\label{eq:lda_B}
    \mathcal{B} \coloneqq \text{span}\{\boldsymbol{\Sigma}^{-1}(\boldsymbol{\mu}_{h} -  \boldsymbol{\mu}_{1}) \mid h = 2, \ldots, H\} \subseteq \mathbb{R}^p.
\end{equation}
Then, under the \textit{LDA} rule, the following lemma establishes that \(\mathcal{B}\) coincides with both the \textit{CS} and \textit{CDS}.

\vspace{1em}
\begin{lemma}\label{lem:lda_cds}
Let \(\mathcal{B}\) be as defined in \eqref{eq:lda_B} and \(\boldsymbol{\Sigma}_{1} = \ldots = \boldsymbol{\Sigma}_{H} = \boldsymbol{\Sigma}\). Then, under model \eqref{eq:model}, \(\mathcal{S}_{Y \mid \mathbf{X}} = \mathcal{S}_{D(Y \mid \mathbf{X})} = \mathcal{B}\). 
\end{lemma}
\vspace{1em}

Both the \textit{QDA} and \textit{LDA} \textit{discriminant subspaces} contain a linear component. However, the linear component \(\mathcal{L}\) of the \textit{discriminant subspace} under the \textit{QDA} model generally differs from the linear component \(\mathcal{B}\) in \textit{LDA}. Although \(\mathcal{L} \neq \mathcal{B}\), the subspaces coincide when the heteroscedastic covariance matrices in the \textit{QDA} model are pooled according to \(\boldsymbol{\Sigma} = \sum_{h = 1}^{H} \pi_{h} \boldsymbol{\Sigma}_{h}\). In that case, \cite{zhang2019} showed that \(\mathcal{L}\) and \(\mathcal{B}\) span the same subspace when combined with \(\mathcal{Q}\), which yields the following identity.

\vspace{1em}
\begin{lemma}\label{lem:L_and_B}
Let \(\mathcal{L}\), \(\mathcal{Q}\), and \(\mathcal{B}\) be as defined in \eqref{eq:qda_L},  \eqref{eq:qda_Q}, and \eqref{eq:lda_B}, respectively. Then, under model \eqref{eq:model}, \(\mathcal{S}_{Y \mid \mathbf{X}} = \mathcal{S}_{D(Y \mid \mathbf{X})} = \mathcal{L} \cup \mathcal{Q} = \mathcal{B} \cup \mathcal{Q}.\)
\end{lemma}
\vspace{1em}

Here, \(\phi_{\text{QDA}}\) and \(\phi_{\text{LDA}}\) characterize supervised classification decision rules that assign an unlabeled observation \(\mathbf{X} \in \mathbb{R}^p\) to a distinct population. While the subspace structure of these rules governs the directions along which class separation occurs, their effectiveness is ultimately quantified by the associated Bayes' error rate, defined as the probability of misclassification under the optimal rule. For example, if we consider the simple case where \(H = 2\) such that \(\pi_{1} = \pi_{2} = 1/2\), then we can easily show (e.g., \cite{johnson2007}) that the \textit{optimal error rate} (\textit{OER}) under the \textit{LDA} decision rule is given by
\begin{equation}\label{eq:lda_oer}
     \Phi\left( -\frac{1}{2} \sqrt{ (\boldsymbol{\mu}_2 - \boldsymbol{\mu}_1)^{\top} \boldsymbol{\Sigma}^{-1} (\boldsymbol{\mu}_2 - \boldsymbol{\mu}_1)} \right),
\end{equation}
where \(\Phi(\cdot)\) is the \textit{cumulative distribution function} (\textit{CDF}) of a standard normal random variable. This expression yields the optimal Bayes' error rate based on the one-dimensional (\textit{1D}) projection of \(\mathbf{X} \in \mathbb{R}^{p}\) onto the direction \(\boldsymbol{\Sigma}^{-1} (\boldsymbol{\mu}_2 - \boldsymbol{\mu}_1)\), which is in \(\mathcal{B}\). Thus, by Lemma \ref{lem:lda_cds}, we have \(\mathcal{B} = \mathcal{S}_{D(Y \mid \mathbf{X})}\) under the \textit{LDA} rule, which confirms that maximum population separation, and consequently the \textit{OER}, is achieved when we project observations onto the \textit{CDS}. Therefore, for any subspace spanned by an arbitrary nonzero vector \(\mathbf{v} \in \mathbb{R}^{p}\), the Bayes' error rate in that subspace will exceed the global \textit{OER} in \eqref{eq:lda_oer} unless \(\mathbf{v}\) spans the same subspace as \(\mathcal{B}\). However, among a set of \(p\) candidate directions \(\mathbf{v}_1, \ldots, \mathbf{v}_p\), we can use the Bayes' \textit{OER} to determine the subspace most aligned with \(\mathcal{S}_{D(Y \mid \mathbf{X})}\). Thus, under model \eqref{eq:model}, we have \(\mathbf{v}^\top_{j}\mathbf{X}|\{Y = h\} \sim \mathcal{N} (\mathbf{v}_{j}^\top\boldsymbol{\mu}_{h}, \mathbf{v}^\top_{j}\boldsymbol{\Sigma}_{h}\mathbf{v}_{j})\), and,  consequently, the \textit{1D} Bayes' \textit{OER} in the subspace defined by \(\text{span}(\mathbf{v}_{j})\) is given by
\begin{equation}\label{eq:1D_oer}
\varphi_{j} \coloneqq
\begin{cases}
\Phi\left( -\dfrac{1}{2}\dfrac{|\mu_{2j}^{\ast} - \mu_{1j}^{\ast}|}{\sigma^{\ast}_{j}} \right), & \sigma_{1j}^{\ast} = \sigma_{2j}^{\ast} = \sigma^{\ast}_{j} \\
\dfrac{1}{2} + \dfrac{1}{2} \Phi\left( \dfrac{\sigma_{1j}^{\ast}(\mu_{2j}^{\ast} - \mu_{1j}^{\ast}) - \sigma_{2j}^{\ast} \tau}{{\sigma_{2j}^{2}}^{\ast} - {\sigma_{1j}^{2}}^{\ast}} \right)
- \dfrac{1}{2} \Phi\left( \dfrac{\sigma_{1j}^{\ast}(\mu_{2j}^{\ast} - \mu_{1j}^{\ast}) + \sigma_{2j}^{\ast} \tau}{{\sigma_{2j}^{2}}^{\ast} - {\sigma_{1j}^{2}}^{\ast}} \right) \\
\quad + \dfrac{1}{2} \Phi\left( \dfrac{\sigma_{2j}^{\ast}(\mu_{2j}^{\ast} - \mu_{1j}^{\ast}) + \sigma_{1j}^{\ast} \tau}{{\sigma_{2j}^{2}}^{\ast} - {\sigma_{1j}^{2}}^{\ast}} \right)
- \dfrac{1}{2} \Phi\left( \dfrac{\sigma_{2j}^{\ast}(\mu_{2j}^{\ast} - \mu_{1j}^{\ast}) - \sigma_{1j}^{\ast} \tau}{{\sigma_{2j}^{2}}^{\ast} - {\sigma_{1j}^{2}}^{\ast}} \right), & \sigma_{2j}^{\ast} > \sigma_{1j}^{\ast},
\end{cases}
\end{equation}
where \(\mu_{hj}^{\ast}\coloneqq\mathbf{v}^{\top}_{j}\boldsymbol{\mu}_{h}\), \({\sigma_{hj}^{2}}^{\ast} \coloneqq \mathbf{v}^{\top}_{j} \boldsymbol{\Sigma}_{h}\mathbf{v}_{j}\),  
and \(\tau \coloneqq \sqrt{(\mu_{2j}^{\ast} - \mu_{1j}^{\ast})^{2} + ({\sigma_{2j}^{2}}^{\ast} - {\sigma_{1j}^{2}}^{\ast})\log\left({\sigma_{2j}^{2}}^{\ast}/{\sigma_{2j}^{2}}^{\ast}\right)}\). When \(\sigma_{1j}^{\ast} = \sigma_{2j}^{\ast}\), \eqref{eq:1D_oer} directly follows from \eqref{eq:lda_oer}, and when \(\sigma_{1j}^{\ast} \neq \sigma_{2j}^{\ast}\), we use the corresponding expression derived by \cite{wu2022}. Note that in \eqref{eq:lda_oer} and \eqref{eq:1D_oer}, we assume equal prior class probabilities for simplicity of presentation. However, this assumption is not essential, and all subsequent theoretical results remain valid with minor modifications.

\subsection{Select SDR Methods}\label{subsec:2.3_SDR}

\begin{singlespace}
\begin{table}[t!]
    \centering
    \caption{Generalized eigenvalue formulations for select \textit{SDR} methods.}
    \begin{tabular}{c|cc}
    \hline
    Method & \(\mathbf{M}\) & \(\mathbf{N}\) \\
    \hline
    \textit{PCA} & \(\boldsymbol{\Sigma}_{\mathbf{X}}\) & \(\mathbf{I}_{p}\) \\
    \textit{SIR} & \(\text{Cov}\left(\mathbb{E}[\mathbf{X} - \mathbb{E}\{\mathbf{X}|Y\}]\right)\) & \(\boldsymbol{\Sigma}_{\mathbf{X}}\) \\
    \textit{SAVE} & \(\Sigma_\mathbf{X}^{1/2} \mathbb{E} \left[ \left\{\mathbf{I}_{p} - \text{Cov}(\mathbf{Z} \mid Y) \right\}^2 \right] \Sigma_\mathbf{X}^{1/2}\) & \(\boldsymbol{\Sigma}_{\mathbf{X}}\) \\
    \textit{SIR-II} & \(\mathbb{E}\left[\text{Cov}(\mathbf{Z}|Y) - \mathbb{E}\{\text{Cov}(\mathbf{Z}|Y)\}\right]^{2}\) & \(\boldsymbol{\Sigma}_{\mathbf{X}}\)\\
    \textit{DR} & \(\boldsymbol{\Sigma}^{1/2}_{\mathbf{X}}\big{\{}[2\mathbb{E}[\mathbb{E}^{2}(\mathbf{ZZ}^{T}\mid Y)] + 2\mathbb{E}^{2}[\mathbb{E}(\mathbf{Z}\mid Y)\mathbb{E}(\mathbf{Z}^{T}\mid Y)] + \) & \(\boldsymbol{\Sigma}_{\mathbf{X}}\) \\
    \(\quad\) & \(2\mathbb{E}\left[\mathbb{E}(\mathbf{Z}\mid Y)\mathbb{E}(\mathbf{Z}\mid Y)\right]\mathbb{E}\left[\mathbb{E}(\mathbf{Z}\mid Y)\mathbb{E}(\mathbf{Z}^{T}\mid Y)\right] - 2\mathbf{I}_{p}]\big{\}}\boldsymbol{\Sigma}^{1/2}_{\mathbf{X}}
\) & \(\quad\) \\
    \textit{SSDR} & \(\left(\mathcal{L}, \boldsymbol\Sigma_2 - \boldsymbol\Sigma_1,\ldots, \boldsymbol\Sigma_H - \boldsymbol\Sigma_1\right)\) \(\left(\mathcal{L}, \boldsymbol\Sigma_2 - \boldsymbol\Sigma_1,\ldots, \boldsymbol\Sigma_H - \boldsymbol\Sigma_1\right)^{\top}\)& \(\mathbf{I}_{p}\)  \\
    \hline
    \end{tabular}
    \label{tbl:SDR}
\end{table}
\end{singlespace}

\cite{li2007} has shown that most \textit{SDR} methods can be formulated as a generalized eigenvalue problem given by
\begin{equation*}
    \mathbf{M} \mathbf{v}_j = \lambda_j \mathbf{N} \mathbf{v}_j, j = 1, \ldots, p,
\end{equation*}
where \(\mathbf{M} \in \mathbb{S}^{p \times p}\) is a method-specific symmetric kernel matrix, and \(\mathbf{N} \in \mathbb{S}^{p}_{+}\) is often taken to be the common covariance matrix, denoted as \(\boldsymbol{\Sigma}_{\mathbf{X}}\). Additionally, \(\mathbf{v}_1, \ldots, \mathbf{v}_p\) are the eigenvectors satisfying \(\mathbf{v}_j^\top \mathbf{N} \mathbf{v}_\ell = 1\) for \(j = \ell\) and \(\mathbf{v}_j^\top \mathbf{N} \mathbf{v}_\ell = 0\) for \(\ell \neq j\), corresponding to the eigenvalues \(\lambda_1 \geq \ldots \geq \lambda_p\). We denote the generalized eigenvalue problem with matrices \(\mathbf{M}\) and \(\mathbf{N}\) as \(\text{GEV}(\mathbf{M}, \mathbf{N})\). Let \(\mathbf{Z} \coloneqq \boldsymbol{\Sigma}_\mathbf{X}^{-1/2} \{\mathbf{X} - \mathbb{E}(\mathbf{X})\}\). Then, we summarize the generalized eigenvalue problem for the following \textit{SDR} methods in Table \ref{tbl:SDR}: \textit{principal components analysis} (\textit{PCA}) introduced by \cite{pearson1901}, \textit{SIR} by \cite{li1991}, \textit{SAVE} by \cite{cook1991}, a variant of \textit{SIR} that considers second-order moments referred to as \textit{sliced inverse regression II} (\textit{SIR-II}) by \cite{li1991}, \textit{directional regression} (\textit{DR}) by \cite{liB2007}, which aims to achieve an exhaustive estimate of the \textit{CS}, and \textit{stabilized sufficient dimension reduction} (\textit{SSDR}) by \cite{boonstra2025}, which adjusts for heteroscedasticity and estimates a similar subspace as the \textit{CDS} under \textit{QDA}. 

Let \(\mathbf{V} = (\mathbf{v}_1, \ldots, \mathbf{v}_p) \in \mathbb{R}^{p \times p}\), with \(\mathbf{V}^\top \mathbf{V} = \mathbf{I}_p\), be the collection of eigenvectors from solving \(\text{GEV}(\mathbf{M}, \mathbf{N})\). In practice, \(\mathbf{M}\) and \(\mathbf{N}\) are replaced by their maximum-likelihood estimates, \(\widehat{\mathbf{M}}\) and \(\widehat{\mathbf{N}}\), respectively, yielding \(\text{GEV}(\widehat{\mathbf{M}}, \widehat{\mathbf{N}})\). Solving this problem yields \(\widehat{\mathbf{V}} = (\widehat{\mathbf{v}}_1, \ldots, \widehat{\mathbf{v}}_p) \in \mathbb{R}^{p \times p}\), with \(\widehat{\mathbf{V}}^{\top}\widehat{\mathbf{V}} = \mathbf{I}_{p}\), the estimated basis obtained from the sample-based \textit{SDR} procedure. Let \(\mathbf{v}_{j}\) and \(\widehat{\mathbf{v}}_{j}\) denote the \(j^{\text{th}}\) vectors of \(\mathbf{V}\) and \(\widehat{\mathbf{V}}\), respectively. Then, under standard moment conditions ensuring \(\widehat{\mathbf{M}}, \widehat{\mathbf{N}} \xrightarrow{P} \mathbf{M}, \mathbf{N}\), and provided the relevant eigenvalues are distinct, \(\widehat{\mathbf{v}}_{j}\) is a root-\(n\) consistent estimator of \(\mathbf{v}_{j}\). This property is commonly established via eigenvector perturbation theory, as detailed in \citet{anderson2003}, and has also been shown by \citet{li1991}, among others, as a property of well-conditioned \textit{SDR} methods.

\section{\small Subspace Ordering Criterion for Binary Response Preservation}\label{sec:3_T}

\subsection{Method}\label{subsec:ill_ex}

To assess the discriminant information of a \textit{DRS}, we consider the independent \textit{Student's T-statistic} introduced by \cite{welch1947},
\begin{equation}\label{eq:t_og}
    T \coloneqq \frac{\overline{x}_{2} - \overline{x}_{1}}{\sqrt{s^{2}_{2}/n_{1} + s^{2}_{1}/n_{2}}}.
\end{equation}

\cite{fan2008} introduced the absolute value of \eqref{eq:t_og} as a now widely used screening tool to select relevant features in high-dimensional data analysis and reduce computational burden (e.g., see \cite{thudumu2020} and \cite{fan2020}). In their theoretical justification of the screening criterion, \cite{fan2008} showed that the absolute value of \eqref{eq:t_og} for the \(j^{\text{th}}\) feature converges in probability to \(\left|\mu_{1j} - \mu_{2j}\right| / \sqrt{\sigma^{2}_{1j}/n_{1} + \sigma^{2}_{2j}/n_{2}}\) under bounded variance and minimal signal assumptions. Although they did not explicitly label this expression, the authors use it throughout their asymptotic arguments (e.g., Theorem 3), and it functions as a population \textit{signal-to-noise ratio} (\textit{SNR}) for predictors. Motivated by their work, we extend the notion of the population \textit{SNR} to a \textit{DRS}.

For each direction \(\mathbf{v}_{j}\), we define the projected \textit{SNR} between populations as 
\begin{equation}\label{eq:Delta_j}
    \Delta_{j} \coloneqq \frac{\left| \mathbf{v}^{\top}_{j}(\boldsymbol{\mu}_{2} - \boldsymbol{\mu}_{1}) \right|}{\sqrt{ \pi_{2}\mathbf{v}^{\top}_{j}\boldsymbol{\Sigma}_{2}\mathbf{v}_{j} + \pi_{1}\mathbf{v}^{\top}_{j}\boldsymbol{\Sigma}_{1}\mathbf{v}_{j}}}, \quad j = 1, \ldots, p.
\end{equation}
To estimate this quantity from data, we define the first two sample moments in the subspace spanned by \(\widehat{\mathbf{v}}_{j}\) as 
\[
\bar{x}^{*}_{hj} \coloneqq \frac{1}{n_h} \sum_{i=1}^{n_h} \widehat{\mathbf{v}}_j^\top \mathbf{x}_{hi}, \quad 
s_{hj}^{2^*} \coloneqq \frac{1}{n_h - 1} \sum_{i=1}^{n_h} \left( \widehat{\mathbf{v}}_j^\top \mathbf{x}_{hi} - \bar{x}^{*}_{hj} \right)^2,
\]
where \(\widehat{\pi}_{h} \coloneqq n_{h} / \sum_{h = 1}^{H} n_{h}\). Thus, we define the sample analogue of the \textit{SNR} in a \textit{DRS} as
\begin{equation}\label{eq:T_j}
    T_j \coloneqq \frac{|\bar{x}^{\ast}_{2j} - \bar{x}^{\ast}_{1j}|}{\sqrt{\widehat{\pi}_{2}s_{2j}^{2\ast} + \widehat{\pi}_{1}s_{1j}^{2\ast}}}, \quad j = 1, \ldots, p.
\end{equation}
Here, \(T_j\) serves as an estimate of \(\Delta_j\) and quantifies the standardized separation between the two class means along the direction \(\widehat{\mathbf{v}}_j\). Thus, we can use \(T_{j}\) to evaluate and rank the components of a \textit{DRS}.
% by their discriminative information.
% As with its feature-wise counterpart, \(T_j\) can be used to 

To illustrate \(T_{j}\) as a subspace criterion, we consider a simple example in which the leading eigenvectors do not necessarily correspond to the subspace that best preserves the relationship between the response and predictors. For this example, we focus on the simple \textit{SDR} method of \textit{PCA}. From Table \ref{tbl:SDR}, \textit{PCA} can be formulated as solving \(\text{GEV}(\boldsymbol{\Sigma}_{X}, \mathbf{I}_{p})\), and the resulting \textit{DRS} is defined by \(\text{span}(\boldsymbol{\Sigma}_{X})\). Consider the covariance structure \(\boldsymbol{\Sigma}_{X} = \text{diag}(3, 2, 1)\), a diagonal matrix with diagonal entries 3, 2, and 1. One can easily show that the eigenvectors of \(\boldsymbol{\Sigma}_{X}\) are \(\mathbf{v}_1 = (1, 0, 0)^\top\), \(\mathbf{v}_2 = (0, 1, 0)^\top\), and \(\mathbf{v}_3 = (0, 0, 1)^\top\), and the respective eigenvalues are \(\lambda_1 = 3\), \(\lambda_2 = 2\), and \(\lambda_3 = 1\). Thus, because \(\lambda_{1} > \lambda_{2} > \lambda_{3}\), the basis for the traditional \textit{DRS} is given by \(\mathbf{V} = (\mathbf{v}_{1}, \mathbf{v}_{2}, \mathbf{v}_{3})\), with \(\mathbf{v}_{1}\) corresponding to the first \textit{DRS} with the largest variance. However, as we will demonstrate, \(\mathbf{v}_{1}\) does not necessarily span the subspace that best captures the relationship between the response and the predictors. In particular, when we consider a binary response, we show that \(T_{j}\) provides a better criterion for maximum class separation.

\begin{figure}[t]
    \centering
    \includegraphics{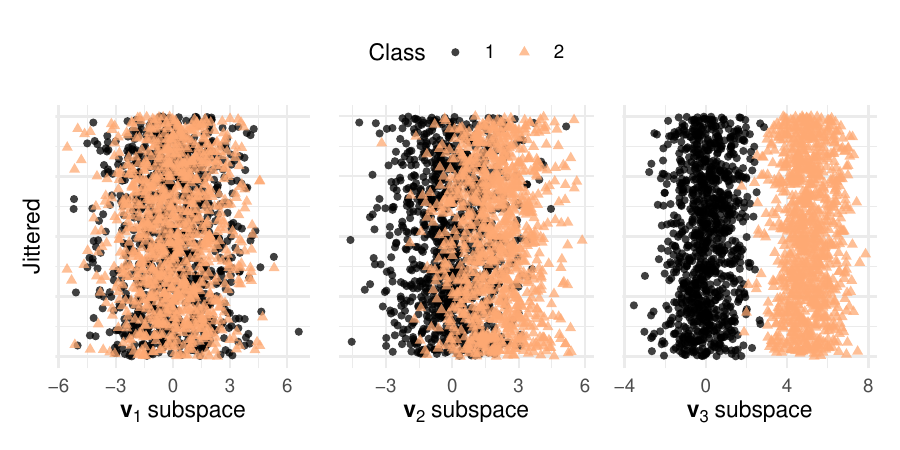}
    \caption{Simulated data from two multivariate normal populations described in Section \ref{subsec:ill_ex} in the subspaces spanned by the leading eigenvectors \(\mathbf{v}_1\), \(\mathbf{v}_{2}\), and \(\mathbf{v}_{3}\). Vertical jitter is added for visualization purposes.}
    \label{fig:1_ill_ex}
\end{figure}

Next, consider the simple parameter configuration \(\boldsymbol{\mu}_{1} = (0, \varepsilon, \alpha)^\top\) and \(\boldsymbol{\mu}_{2} = (0, 0, 0)^\top\), where \(0 < |\varepsilon| \ll |\alpha|\), and assume both populations share the common covariance matrix \(\boldsymbol{\Sigma}_{X} = \text{diag}(3, 2, 1)\). Clearly, the discriminant information lies almost entirely in the third coordinate through \(\alpha\) while the contributions from the first and second coordinates are negligible for a relatively small \(\varepsilon\). Thus, only \(\mathbf{v}_{3} = (0, 0, 1)^{\top}\), the eigenvector with the smallest eigenvalue, contains the dominant signal. This result is captured in the corresponding \(\Delta_j\) values for each subspace spanned by \(\mathbf{v}_j\) that  simplify to \(\Delta_{j} = \left| \mathbf{v}^\top_{j}\boldsymbol{\mu}_{1} \right| / \sqrt{\lambda_{j}}\) because \(\lambda_{j} = \mathbf{v}_{j}^{\top}\boldsymbol{\Sigma}\mathbf{v}_{j}\) in the \textit{PCA} setting. Hence, for \(\mathbf{v}_{1}\), \(\mathbf{v}_{2}\), and \(\mathbf{v}_{3}\), we have \(\Delta_{1} = 0\), \(\Delta_{2} = |\varepsilon|/\sqrt{2}\), and \(\Delta_{3} = |\alpha|\), respectively. Since \(|\alpha| > |\varepsilon|\), we have \(\Delta_{3} > \Delta_{2} > \Delta_{1}\), and in the event that \(\varepsilon \approx \alpha\), we still have \(\Delta_{3} > \Delta_{2}\) because \(\lambda_{2} > \lambda_{3}\). Thus, \(\Delta_{j}\) indicates that \(\mathbf{v}_{3}\), not \(\mathbf{v}_{1}\), defines the most important subspace in terms of classification. Therefore, we should reorder the \textit{DRS} as \(\mathbf{V} = (\mathbf{v}_{3}, \mathbf{v}_{2}, \mathbf{v}_{1})\).

To visualize this example, we generated 1,000 observations from each population, in which we took \(\alpha = 5\) and \(\varepsilon = 2\) and assumed each population followed a multivariate normal distribution. In Figure \ref{fig:1_ill_ex}, the simulated data is given in each subspace defined by \(\mathbf{v}_{1}\), \(\mathbf{v}_{2}\), and \(\mathbf{v}_{3}\), respectively. Clearly, superior class separation was achieved in \(\mathbf{v}_{3}\) compared to \(\mathbf{v}_{2}\) or \(\mathbf{v}_{1}\). Moreover, this ordering was preserved in the estimated basis \(\widehat{\mathbf{V}} = (\widehat{\mathbf{v}}_{1}, \widehat{\mathbf{v}}_{2}, \widehat{\mathbf{v}}_{3})\), computed from the simulated data, where the sample principal components were determined to be \(\widehat{\mathbf{v}}_{1} = (1.00, 0.01, -0.01)^{\top}\), \(\widehat{\mathbf{v}}_{2} = (-0.01, 1.00, -0.01)^{\top}\), and \(\widehat{\mathbf{v}}_{3} = (0.01, 0.01, 1.00)^{\top}\). The corresponding eigenvalues were \(\widehat{\lambda}_{1} = 3.23\), \(\widehat{\lambda}_{2} = 2.06\), and \(\widehat{\lambda}_{3} = 1.01\). Under a traditional eigenvalue or variance-based approach, \(\widehat{\mathbf{v}}_1\) would be prioritized as the most informative subspace. However, the \textit{SNR}s computed via \(T_j\) tell a different story. We found \(T_1 = 0.003\), \(T_2 = 1.36\), and \(T_3 = 4.97\), clearly indicating that \(\widehat{\mathbf{v}}_3\) was the most discriminative subspace. Consistent with this finding, the estimated Bayes’ error rates under the \textit{LDA} decision rule for the one-dimensional subspaces spanned by \(\widehat{\mathbf{v}}_{1}\), \(\widehat{\mathbf{v}}_{2}\), and \(\widehat{\mathbf{v}}_{3}\) were 0.5040, 0.2560, and 0.0045, respectively. Thus, \(\widehat{\mathbf{v}}_3\), the eigenvector corresponding to the smallest eigenvalue and traditionally considered the least informative under \textit{PCA}, was in fact the most important subspace for discrimination. Therefore, using \(T_j\) rather than eigenvalue magnitude as a subspace ordering criterion corrects a fundamental misalignment in \textit{SDR} for supervised learning by prioritizing subspaces that preserve the predictor–response relationship rather than the subspaces that merely aim to maximize overall variability. We formalize this criterion in the next section with theoretical results.

\subsection{Theoretical Properties}\label{sec:3.1_T_theory}

Here, we study the theoretical properties of \(\Delta_j\) as an importance measure for a \textit{DRS} by establishing its connection to the \textit{CDS} and the Bayes' error rate. We then establish the consistency of the sample estimate \(T_j\) by showing that the ranking induced by the \(T_j\)s converges to that of the \(\Delta_j\)s under mild regularity conditions. We begin by providing the theorem below, which states the necessary conditions under which \(\Delta_{j}\) can identify whether a subspace is aligned with the \textit{CDS}.

\vspace{1em}
\begin{thm}\label{thm:T_CDS}
    Let \(\mathbf{X} \in \mathbb{R}^p \) and \( Y \in \{1, 2\} \) follow model \eqref{eq:model}, \(\boldsymbol{\Sigma} = \sum_{h = 1}^{2}\pi_{h}\boldsymbol{\Sigma_{h}}\), and \(\Delta_{j}\) be as defined in \eqref{eq:Delta_j}. If \( \Delta_j \neq 0 \), then \( \boldsymbol{\Sigma}\mathbf{v}_j \not\perp \mathcal{S}_{D(Y \mid \mathbf{X})} \).  
\end{thm}
\vspace{1em}

Theorem \ref{thm:T_CDS} is stated in terms of \(\boldsymbol{\Sigma} \mathbf{v}_{j}\) rather than for any arbitrary \(\mathbf{v}_{j}\) itself because, under model \eqref{eq:model}, the \textit{CDS} for both \textit{LDA} and \textit{QDA} is characterized by the span of covariance-adjusted mean difference vectors of the form \(\boldsymbol{\Sigma}^{-1}(\boldsymbol{\mu}_h - \boldsymbol{\mu}_1)\) in \(\mathcal{B}\). Thus, to determine whether a candidate direction is aligned with the \textit{CDS}, we naturally  assess \(\boldsymbol{\Sigma}\mathbf{v}_j\), which places the direction in the same covariance-adjusted subspace as \(\mathcal{S}_{D(Y \mid X)}\). Moreover, in the \(\text{GEV}(\mathbf{M}, \mathbf{N})\) formulation for an \textit{SDR} basis when \(\mathbf{N} = \boldsymbol{\Sigma}\), which is often the case, the relevant population subspace is likewise expressed through \(\boldsymbol{\Sigma} \mathbf{v}_{j}\).

Thus, Theorem \ref{thm:T_CDS} establishes that \(\Delta_j\) provides a criterion for determining whether the covariance-adjusted subspace spanned by a candidate direction \(\mathbf{v}_j\) at least partially recovers \(\mathcal{S}_{D(Y \mid X)}\). That is, \(\Delta_j\) yields a measure for detecting whether any given subspace carries discriminatory signal. However, in general, many of the \(p\) candidate directions produced by an \textit{SDR} method may at least partially estimate the \textit{CDS}. Therefore, we further establish \(\Delta_{j}\) as a subspace criterion by demonstrating in the theorem below that any candidate direction with a larger \(\Delta_{j}\) will yield a smaller Bayes' error rate.

\vspace{1em}
\begin{thm}\label{thm:T_Bayes}
Let \(\mathbf{X} \in \mathbb{R}^p \) and \( Y \in \{1, 2\} \) follow model \eqref{eq:model}. Let \( \mathbf{v}_j, \mathbf{v}_\ell\) exist such that \(\mathbf{v}_j \neq \mathbf{v}_\ell\). Let \(\varphi_{j}\) and \(\Delta_{j}\)  be as defined in \eqref{eq:1D_oer} and \eqref{eq:Delta_j}, respectively. 
Then, we have the following results.
\begin{enumerate}
    \item Let \(\boldsymbol{\Sigma}_{1} = \boldsymbol{\Sigma}_{2}\). Then, \(\Delta_{j} > \Delta_{\ell}\) if and only if \(\varphi_{j} < \varphi_{\ell}\). 
    \item Let \(\boldsymbol{\Sigma}_{1} \neq \boldsymbol{\Sigma}_{2}\), and suppose  \(\mathbf{v}_{j}^{\top}\boldsymbol{\Sigma}_{2}\mathbf{v}_{j} =\mathbf{v}_{\ell}^{\top}\boldsymbol{\Sigma}_{2}\mathbf{v}_{\ell}\). Then, \(\Delta_{j} > \Delta_{\ell}\) if and only if \(\varphi_{j} < \varphi_{\ell}\).
\end{enumerate}
\end{thm}
\vspace{1em}

Although \(\Delta_{j}\) allows \(\boldsymbol{\Sigma}_{1} \neq \boldsymbol{\Sigma}_{2}\), it does not directly measure heteroscedasticity. Thus, for Theorem \ref{thm:T_Bayes} under the \textit{QDA} model, we impose that \(\mathbf{v}_{j}^{\top}\boldsymbol{\Sigma}_{2}\mathbf{v}_{j} = \mathbf{v}_{\ell}^{\top}\boldsymbol{\Sigma}_{2}\mathbf{v}_{\ell}\). That is, for one population only, the variability in the subspaces spanned by \(\mathbf{v}_{j}\) and \(\mathbf{v}_{\ell}\) is similar. Under this imposed constraint, \(\varphi_{j}\) is a monotonic function of \(\Delta_{j}\). This assumption is easily satisfied, but not limited to, when either population covariance matrix is spherical. No additional assumptions are required under the \textit{LDA} model. 

By Theorem \ref{thm:T_Bayes}, we can directly compare the discriminatory strength of two subspaces using \(\Delta_{j}\) under both the \textit{QDA} and \textit{LDA} decision rules. This result formalizes a key implication: Among a set of subspaces, the direction \(\mathbf{v}_{j}\) with the largest \(\Delta_j\) achieves the local minimum Bayes' error rate. When \(H = 2\) under \textit{LDA}, Lemma \ref{lem:lda_cds} yields \(d = 1\). Thus, we have that selecting the top-ranked direction by \(\Delta_{j}\) directly estimates \(\mathcal{S}_{D(Y\mid \mathbf{X})}\) and yields the local minimum Bayes' error rate. Under \textit{QDA}, Lemma \ref{lem:qda_cds} yields \(d = \text{rank}(\mathcal{L} \cup \mathcal{Q}) \geq 1\). Thus, while ordering by \(\Delta_j\) still identifies directions with minimal \textit{1D} Bayes' error rates, the \(d\)-dimensional error rate cannot be determined from individual rankings due to the lack of a tractable \textit{OER} when \(p > 1\). However, as shown in the simulation studies in Section \ref{sec:5_simulation}, combining top-ranked directions consistently yields substantial reductions in the estimated Bayes' error rates, which demonstrates the practical gain of reordering subspaces under \textit{QDA}. Therefore, we can reorder the subspaces resulting from an \textit{SDR} method  by their corresponding \(\Delta_j\) values to maximize population separation. 

The previous theoretical results establish \(\Delta_j\) as an importance measure for evaluating and ordering directions in a \textit{DRS}. Now we establish the consistency of its sample analogue \(T_j\). Our interest is not solely in the pointwise convergence of each statistic but rather in the behavior of the entire vector \((T_{1}, \ldots, T_{p})^{\top}\) and its properties as a ranking criterion. Specifically, for any vector \(\boldsymbol{\theta} = (\theta_1, \ldots, \theta_p)^{\top} \in \mathbb{R}^p\) and \(j \in \{1, \ldots, p\}\), we define the \textit{rank-order} vector as
\begin{equation}\label{eq:R}
    \mathcal{R}(\boldsymbol{\theta}) \coloneqq (r_1(\boldsymbol{\theta}), \ldots, r_p(\boldsymbol{\theta}))^{\top}, \text{ where } r_j \coloneqq \sum_{i=1}^p \mathbf{1}\{ \theta_i > \theta_j \} + 1.
\end{equation}
Here, \(r_{j}(\boldsymbol{\theta})\) denotes the relative \textit{rank-order} of the \(j^{\text{th}}\) component of \(\boldsymbol{\theta}\), which assigns the lowest value of 1 to the largest entry. We first provide the necessary conditions under which uniform convergence of a finite-dimensional vector of estimators guarantees consistency of the induced rank-order in the lemma below. 

\vspace{1em}
\begin{lemma}\label{lemma:rank}
Let $\boldsymbol{\theta} \coloneqq (\theta_1, \ldots, \theta_p)^{\top} \in \mathbb{R}^{p}$ be fixed, and let $\hat{\boldsymbol{\theta}} \coloneqq (\hat{\theta}_1^{(n)}, \ldots, \hat{\theta}_p^{(n)})^{\top} \in \mathbb{R}^{p}, n \in \mathbb{N}$. Suppose that \(\hat{\theta}_j^{(n)} \xrightarrow{P} \theta_j,\ j = 1, \ldots, p\), and there exist an  \(\varepsilon > 0\) such that
\(\left|\ \theta_i - \theta_j \,\right| \ge \varepsilon, \text{ for all }  i \ne j\). Let \(\mathcal{R}(\cdot)\) be as defined in \eqref{eq:R}.
Then, $\mathbb{P}\left( \mathcal{R}(\hat{\boldsymbol{\theta}}) = \mathcal{R}(\boldsymbol{\theta}) \right) \to 1$.
\end{lemma}
\vspace{1em}

Fundamental to our results in the theorems that follow, this lemma applies beyond the specific case of consistency for subspace ordering induced by the \(T_{j}\)s. The result requires that the entries of the population vector \(\boldsymbol{\theta}\) be unique. We note that this condition is not restrictive because most ranking procedures assume distinct entries or break ties arbitrarily. Moreover, in our context, \(\Delta_j\) is a continuous function of the model parameters and \(\Delta_{j} = 0\) if either \(\boldsymbol{\mu}_1 = \boldsymbol{\mu}_2\) (i.e., no signal exists) or \(\mathbf{v}_{j} \perp \mathcal{S}_{D(Y \mid \mathbf{X})}\). To establish the \textit{rank-order} consistency of the \(T_{j}\)s, we work under relaxed conditions with no assumption of multivariate normality. We impose only the following conditions. 

\noindent \textbf{Conditions.}
\begin{itshape}
\begin{enumerate}
    \item[(C1)] We have \(\mathbf{X}_{hi} = \boldsymbol{\mu}_h + \boldsymbol{\varepsilon}_{hi}\) for \(h = 1, \ldots, H\) and \(i = 1, \ldots, n_h\). 
    \item[(C2)] The vectors \(\boldsymbol{\varepsilon}_{hi}\) are IID within each population \(h\) such that \(\mathbb{E}(\boldsymbol{\varepsilon}_{hi}) = \mathbf{0}\) and \(\text{Cov}(\boldsymbol{\varepsilon}_{hi}) = \boldsymbol{\Sigma}_h \in \mathbb{S}_+^p\). Additionally, \(\boldsymbol{\varepsilon}_{h\cdot}\) are independent across populations.  
    \item[(C3)] Each component of \(\boldsymbol{\varepsilon}_{hi}\) has a finite second moment.  
\end{enumerate}
\end{itshape}

\vspace{1em}
\begin{thm}\label{thm:T_rank} 
    Suppose conditions C1 - C3 hold, and let \(\Delta_{j}\) and \(T_{j}\), \(j = 1, \ldots, p\), be as defined in \eqref{eq:Delta_j} and  \eqref{eq:T_j}, respectively. Let \(\mathcal{R}(\cdot)\) be defined as in \eqref{eq:R}. Then, we have \(\mathbb{P}\left( \mathcal{R}(T_{1}, \ldots, T_{p}) = \mathcal{R}(\Delta_{1}, \ldots, \Delta_{p}) \right) \to 1\). 
\end{thm}
\vspace{1em}

Theorem \ref{thm:T_rank} formalizes the \textit{rank-order} consistency of the sample-based \(T_j\)s with respect to the population-level \(\Delta_j\)s. This result relies on the fact that the estimated eigenvectors are root-\(n\) consistent, as discussed in Section \ref{subsec:2.3_SDR}. This theorem ensures that, asymptotically, the ordering of subspaces by their empirical \(T_j\) values recovers the population ordering induced by the \(\Delta_j\)s. Therefore, Theorems \ref{thm:T_Bayes} and \ref{thm:T_rank} establish that \(T_{j}\) provides a consistent measure for evaluating and ordering directions in a \textit{DRS}.

\section{\large Subspace Criterion for a Categorical or Continuous Response}\label{sec:4_F}

In the previous section, we established the use of \(T_{j}\) as a criterion for ordering subspaces when the response is binary. When the response is categorical with \(H \geq 2\), we extend this subspace ordering criterion by proposing an \(F\)-statistic, introduced by \cite{fisher1925}, within a \textit{DRS} as an importance measure for subspaces, which we define as 
\begin{equation}\label{eq:F_j}
F_j \coloneqq
\frac{\sum_{h=1}^H \widehat{\pi}_{h} \left( \bar{x}^{\ast}_{hj} - \sum_{i = 1}^{H}\widehat{\pi}_{i}\bar{x}^{\ast}_{ij} \right)^2}
     {\sum_{h=1}^H \widehat{\pi}_{h} s_{hj}^{2\ast}},
\quad  j = 1, \ldots, p.
\end{equation}
Moreover, the novelty of the \(F_{j}\) measure lies in its ability to unify categorical and continuous responses under a single subspace criterion. Under the slicing-based framework in \textit{SDR}, we can use the \(H\) slices of \(Y\) as categories in which the \(F_{j}\) criterion can also be applied to continuous responses. Thus, for either a categorical or a continuous response, the \(F_{j}\) criterion provides a measure of between-slice separation relative to within-slice variability for each subspace. For a categorical response, this criterion identifies subspaces that maximize overall population separation whereas, for a continuous response, ordering by \(F_{j}\) yields subspaces that maximize slice separation, with the goal of identifying directions that preserve the conditional mean structure as the number of slices increases.

When the response \(Y\) is discretized, either naturally through predefined populations or artificially through slicing, such that \(\widetilde{Y} \in \{1, \ldots, H\}\) with \(H \geq 2\), we formalize \(F_{j}\) as a subspace criterion by studying the theoretical properties of its population analogue,
\begin{equation}\label{eq:Psi_j}
\Psi_{j} \coloneqq
\frac{
\displaystyle \sum_{h=1}^H \pi_h \left( \mathbf{v}_j^\top \boldsymbol{\mu}_h - \sum_{i=1}^H \pi_i\, \mathbf{v}_j^\top \boldsymbol{\mu}_i \right)^2
}{
\displaystyle \sum_{h=1}^H \pi_h \, \mathbf{v}_j^\top \boldsymbol{\Sigma}_h \mathbf{v}_j
}, \quad j = 1, \ldots, p.
\end{equation}
If \(H = 2\), then Corollary \ref{cor:T_and_F} below formalizes the natural extension from \(T_{j}\) to \(F_{j}\) by establishing that \(\Psi_{j}\) yields an identical ordering of subspaces as \(\Delta_{j}\) and, as a result, retains the same theoretical properties for discrimination as \(\Delta_{j}\).

\vspace{1em}
\begin{cor}\label{cor:T_and_F}
Let H = 2. For each \(\mathbf{v}_j\), \(j = 1, \ldots, p\), let \(\Delta_{j}\) and \(\Psi_{j}\) be as defined in \eqref{eq:Delta_j} and \eqref{eq:Psi_j}, respectively. Let \(\mathcal{R}(\cdot)\) be as defined in \eqref{eq:R}. Then, \(\Delta_{j} \propto \sqrt{\Psi}_{j}\) and \(\mathcal{R}(\Psi_{1}, \ldots, \Psi_{p}) = \mathcal{R}(\Delta_{1}, \ldots, \Delta_{p})\). 
\end{cor}
\vspace{1em}

In Theorem \ref{thm:T_CDS}, we established that when the response is binary, \(\Delta_j\) identifies directions that are at least partially aligned with \(\mathcal{S}_{D(Y \mid \mathbf{X})}\). The following theorem shows that \(\Psi_j\) retains this property in the multiclass setting and that for a continuous response we obtain a similar result in terms of \(\mathcal{S}_{Y|\mathbf{X}}\).

\vspace{1em}
\begin{thm}\label{thm:F_CDS_CS}
    Let \(\mathbf{X} \in \mathbb{R}^p \) and \( \widetilde{Y} \in \{1, \ldots, H\} \). Let \(\boldsymbol{\Sigma} = \sum_{h = 1}^{H}\pi_{h}\boldsymbol{\Sigma_{h}}\) and \(\Psi_{j}\) be as defined in \eqref{eq:Psi_j}. We then have the following results. 
    \begin{enumerate}
        \item  Suppose \(\mathbb{E}[\mathbf{X}|\boldsymbol{\beta}^{\top}\mathbf{X}]\) is a linear function  of \(\boldsymbol{\beta}^{\top}\mathbf{X}\) such that \(\text{span}(\boldsymbol{\beta}) = \mathcal{S}_{Y|\mathbf{X}}\) and \(\boldsymbol{\beta} \in \mathbb{R}^{p \times d}\). If \( \Psi_j \neq 0 \), then \( \boldsymbol{\Sigma}\mathbf{v}_j \not\perp \mathcal{S}_{Y \mid \mathbf{X}} \).
        \item Suppose \(\mathbf{X}|\widetilde{Y}\) follows model \eqref{eq:model}. If \( \Psi_j \neq 0 \),  then \( \boldsymbol{\Sigma}\mathbf{v}_j \not\perp \mathcal{S}_{D(Y \mid \mathbf{X})} \). 
    \end{enumerate}
\end{thm}
\vspace{1em}

Theorem \ref{thm:F_CDS_CS} establishes that \(\Psi_{j}\) provides a criterion for detecting informative subspaces that guarantees partial alignment with \(\mathcal{S}_{Y|\mathbf{X}}\) under the linearity condition or with \(\mathcal{S}_{D(Y|\mathbf{X})}\) under model \eqref{eq:model}. For the continuous response case, the linearity condition is a mild and standard assumption in \textit{SDR} that is also satisfied whenever \(\mathbf{X}\) follows an elliptical distribution. To extend these results to the sample setting, we establish that \(F_{j}\) yields a consistent ranking of directions relative to \(\Psi_{j}\), as formalized in Theorem \ref{thm:F_rank}.

\vspace{1em}
\begin{thm}\label{thm:F_rank} 
    Suppose conditions C1 - C3 hold, and let \(F_{j}\) and \(\Psi_{j}\), \(j = 1, \ldots, p\), be as defined in \eqref{eq:F_j} and  \eqref{eq:Psi_j}, respectively. Let \(\mathcal{R}(\cdot)\) be defined as in \eqref{eq:R}. Then, we have \(\mathbb{P}\left( \mathcal{R}(F_{1}, \ldots, F_{p}) = \mathcal{R}(\Psi_{1}, \ldots, \Psi_{p}) \right) \to 1\). 
\end{thm}
\vspace{1em}

Theorems \ref{thm:F_CDS_CS} and \ref{thm:F_rank} establish \(F_{j}\) as a consistent sample-based subspace criterion that preserves the ranking induced by \(\Psi_{j}\), thereby providing a measure for evaluating the relevance of candidate directions. We note that, in contrast to the binary case where \(\Delta_{j}\) yields a direct monotonic relationship with the Bayes' error rate, such a result is not available for \(\Psi_{j}\) because no general tractable \textit{OER} for \textit{LDA} or \textit{QDA} exists when \(H > 2\). However, our simulation results and real-data applications demonstrate the practical effectiveness of ordering subspaces by \(F_{j}\), which identifies directions with greater overall class separation and, as a result, often significantly reduces the estimated misclassification rate. 

For a continuous response, a similar limitation arises because we cannot evaluate any population-level measure of optimality against \(\Psi_{j}\). However, in the sample case, when \(\text{span}(\boldsymbol{\beta}) = \mathcal{S}_{Y|\mathbf{X}}\) must be estimated by \(\widehat{\boldsymbol{\beta}}\), which corresponds to the \textit{SDR} method-specific eigenvectors \(\widehat{\mathbf{V}}\), the distance between the estimated and true subspaces can be quantified by 
\begin{equation}\label{eq:D}
    \mathcal{D}(\mathcal{S}_{\boldsymbol{\beta}}, \mathcal{S}_{\hat{\boldsymbol{\beta}}}) = \frac{{\|\mathbf{P}_{\boldsymbol{\beta}} - \mathbf{P}_{\hat{\boldsymbol{\beta}}}\|}_{F}}{\sqrt{2d}}.
\end{equation}
The subspace distance in both \eqref{eq:D} and similar measures is widely used in \textit{SDR} (e.g., see \cite{cook_and_zhang2014}, \cite{lin2019}, and \cite{zeng2024}) because full-rank rotations of \(\boldsymbol{\beta}\) and \(\widehat{\boldsymbol{\beta}}\) do not change its value. That is, \(\mathcal{D}\) is a coordinate-free measure that ranges between \([0, 1]\) when the estimated and true subspaces have the same dimension. Values closer to 0 indicate that \(\text{span}(\widehat{\boldsymbol{\beta}})\) and \(\mathcal{S}_{Y|\mathbf{X}}\) are closely aligned. Thus, in establishing \(F_{j}\) for a continuous response, we rely on Theorems \ref{thm:F_CDS_CS} and \ref{thm:F_rank} and our empirical results, which demonstrate that ordering a \textit{DRS} by \(F_{j}\) consistently yields smaller values of \(\mathcal{D}\) and indicates that the criterion can often identify directions that better recover \(\mathcal{S}_{Y|\mathbf{X}}\).

\section{Simulation Studies}\label{sec:5_simulation}
\subsection{Simulation for Binary Response}\label{subsec:5.1_CER_sims}
We used \textit{Monte Carlo} (\textit{MC}) simulations to demonstrate the efficacy of reordering the \textit{DRS} of an \textit{SDR} method using the \(T_{j}\) criterion contrasted to the eigenvalue magnitude. The \textit{PCA}, \textit{SAVE}, \textit{SIR-II}, and \textit{SSDR} methods were implemented as described in Section \ref{subsec:2.3_SDR}. Note that we implemented the \textit{SIR-II} method rather than the classical \textit{SIR} method because \textit{SIR} relies only on first-order moments. This fact makes its eigenvalues proportional to \(T_{j}\) and thus yields the same ordering. Using the \(F_{j}\) criterion, we provide additional simulations in the Supplementary Material for categorical responses with \(H > 2\). We refer to the eigenvalue-ordered \textit{SDR} methods by their standard names and the \(T_{j}\)-ordered versions as \(PCA_{T}\), \(SAVE_{T}\), \(SIR\text{-}II_{T}\), and \(SSDR_{T}\), respectively.

We considered parameter configurations from multivariate normal populations that are commonly used for accessing discriminant analysis and similar to those in \cite{mai2019}, \cite{gaynanova2019}, and \cite{boonstra2025}. We set \(p = 50\) for the \textit{MC} simulation study. For configurations Q1-Q3 below, we used \textit{QDA} as the supervised classifier.
\begin{itemize}
    \item Configuration Q1: \(\boldsymbol{\mu}_{1} = \mathbf{0}_{p}\), \(\boldsymbol{\mu}_{2}\) is a \(p \times 1 \) vector with \textit{IID} entries from the \(\mathcal{N}(0, 1)\) distribution, \(\boldsymbol{\Sigma}_{1} = \mathbf{I}_{p}\), \(\boldsymbol{\Sigma}_{2} = \begin{bsmallmatrix} \phantom{-}3 & -2 \\ -2 & \phantom{-}3\end{bsmallmatrix} \oplus \mathbf{I}_{p-2}\), and \(d = 2\). 
    \item Configuration Q2: \(\boldsymbol{\mu}_{1} = \mathbf{0}_{p}\), \(\boldsymbol{\mu}_{2} = (\mathbf{1}_{5}, -\mathbf{1}_{5}, \mathbf{0}_{p - 10})\), \(\boldsymbol{\Sigma}_{1} = \mathbf{I}_{p}\),  and 
    \(\boldsymbol{\Sigma}_{2} = [\rho \mathbf{I}_{b} + (1-\rho)(\mathbf{J}_{b}-\mathbf{I}_{b})] \oplus \mathbf{I}_{p-b} \),  where \(\mathbf{J}_{p}\) is a \(p \times p\) matrix of ones, \(b = 5\), \(\rho = 0.99\), and \(d = 6\). 
    \item Configuration Q3: \(\boldsymbol{\mu}_{1} = \mathbf{0}_{p}\), \(\boldsymbol{\mu}_{2} = (\mathbf{0}_{p - 1}, 1)^{\top}\), \(\boldsymbol{\Sigma}_{1} = \boldsymbol{\Sigma}_{2} = (\sigma_{ij})\), where \(\sigma_{ii} = 2, \text{ for all } i \neq p\), \(\sigma_{pp} = 1\), \(\sigma_{ij} = 0, \text{ for all } i \neq j\),  and \(d = 1\). 
\end{itemize}
We used \textit{LDA} as the supervised classifier for configurations L1-L3 below.
\begin{itemize}
    \item Configuration L1: \(\boldsymbol{\mu}_{1} = \mathbf{0}_{p}\), \(\boldsymbol{\mu}_{2} = \mathbf{1}_{p}\), \(\boldsymbol{\Sigma}_{1} = \boldsymbol{\Sigma}_{2} = (1 - \rho)\mathbf{I}_{p} + \rho\mathbf{J}_{p} \), \(\rho  = 0.25\), and \(d = 1\). 
    \item Configuration L2: \(\boldsymbol{\mu}_{1} = (1, 1, \mathbf{0}_{p - 2})^{\top}\), \(\boldsymbol{\mu}_{2} = - \boldsymbol{\mu}_{1}\), and 
\(\boldsymbol{\Sigma}_{1} = \boldsymbol{\Sigma}_{2} = \mathbf{I}_{2} \oplus s^{2}[(1-\rho)\mathbf{I}_{p-2}+\rho \mathbf{J}_{p-2})]\),  where \(s^{2} =  10\), \(\rho = 0.99\), and \(d = 1\). 
    \item Configuration L3: The same setting as configuration Q2 except \(b = 20\).  Thus, \(d = 20\). 
\end{itemize}

\begin{figure}[t!]
    \centering
    \includegraphics{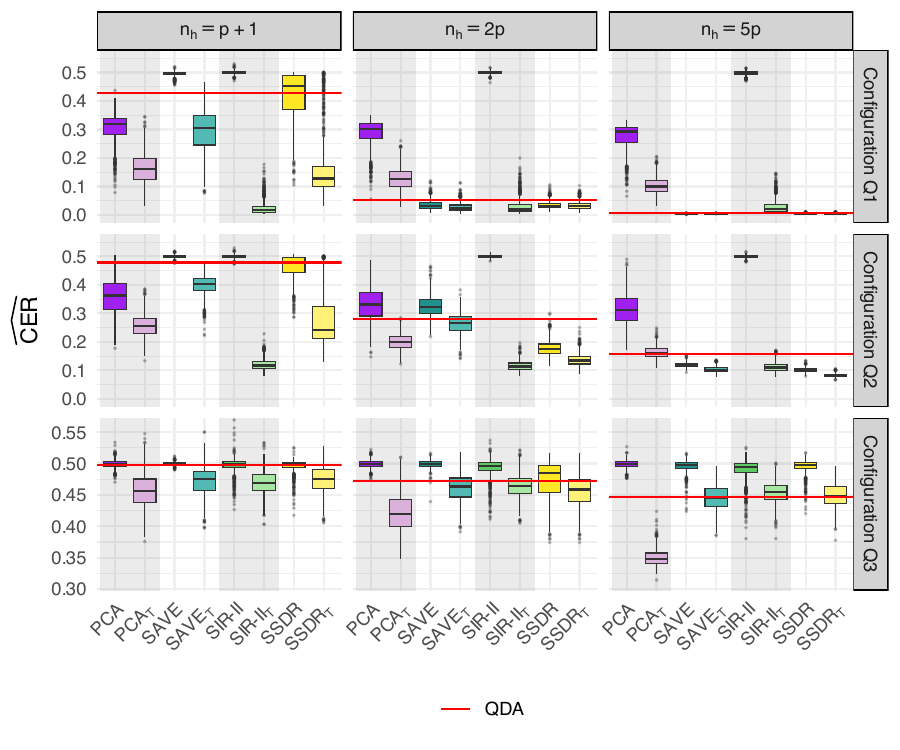}
    \caption{Estimated conditional error rate \((\widehat{CER})\) plots for the simulation described in Section \ref{subsec:5.1_CER_sims} for contrasting eigenvalue-ordered \textit{SDR} methods (labeled by their standard names) with their \(T_{j}\)-ordered counterparts (denoted by the superscript T). All \textit{SDR} methods used the \textit{QDA} classifier, and the horizontal line represents the median
     \(\widehat{CER}\) using \textit{QDA} without dimension reduction.}
    \label{fig:2_QDA}
\end{figure}

\begin{figure}[t!]
    \centering
    \includegraphics{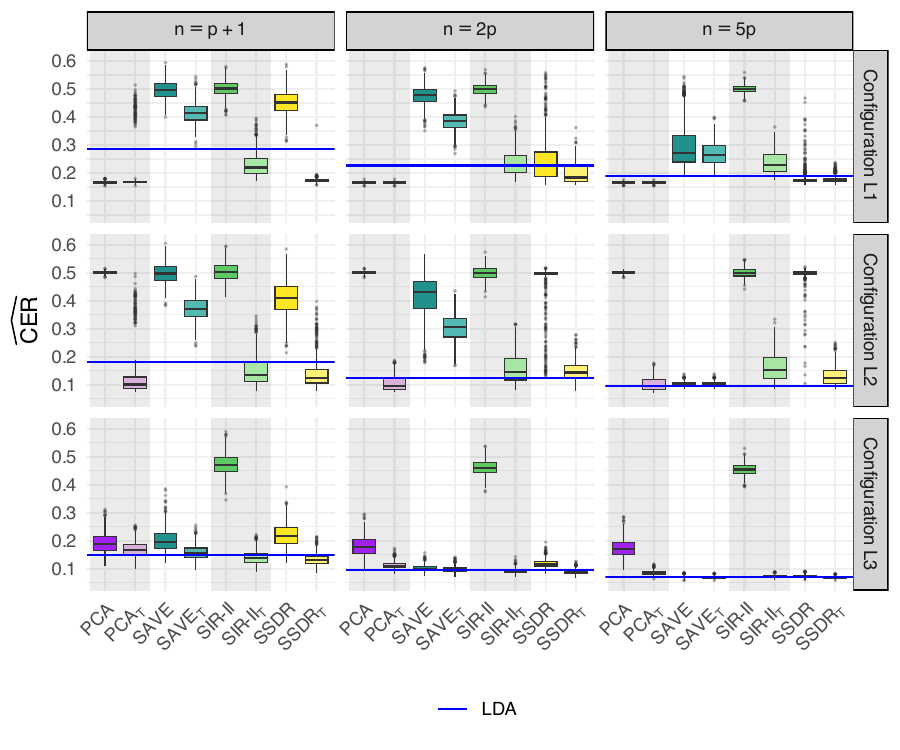}
    \caption{Estimated conditional error rate \((\widehat{CER})\) plots for the simulation described in Section \ref{subsec:5.1_CER_sims}, contrasting eigenvalue-ordered \textit{SDR} methods (labeled by their standard names) with their \(T_{j}\)-ordered counterparts (denoted by the superscript \(T\)). LDA was used as the supervised classifier for all \textit{SDR} methods. The median \(\widehat{CER}\) using \textit{LDA} without dimension reduction is represented by the horizontal line.}
    \label{fig:3_LDA}
\end{figure}

For each configuration, we generated 5,000 observations from each population. We varied the training sample sizes to simulate ill-conditioned to well-conditioned estimation of the discriminant parameters. The class-specific training sample sizes for the \textit{QDA} configurations were \(n_{h} = p + 1,\ 2p,\ \text{and } 5p,\ h \in \{1,2\}\). For the \textit{LDA} configurations, the training sample sizes were \(n = p + 1,\ 2p,\ \text{and } 5p\) with equal prior class probabilities. We used the remaining observations as the test set. For each \textit{SDR} method, we projected the training and test sets from \(p\) to \(d = \text{dim}(\mathcal{S}_{D(Y|\mathbf{X})})\) dimensions, then applied the respective classifier and recorded the estimated \textit{conditional error rate}, denoted by \(\widehat{CER}\). We also recorded the \(\widehat{CER}\) of the full-feature data using the respective classifier without dimension reduction. This process was replicated 1,000 times for each configuration. We summarized the \textit{MC} simulation results in Figures \ref{fig:2_QDA} and \ref{fig:3_LDA} for the \textit{QDA} and \textit{LDA} configurations, respectively, by displaying the distributions of the \(\widehat{CER}\)s for each \textit{SDR} method. The horizontal line denotes the median \(\widehat{CER}\) of the full-dimensional classifier with no dimension reduction.

From Figure \ref{fig:2_QDA}, we clearly see that, compared to ordering by eigenvalue magnitude, ordering the \textit{DRS} by \(T_{j}\) can significantly reduce the \(\widehat{CER}\). In most cases, we found that eigenvalue ordering often yielded \textit{SDR} methods with \(\widehat{CER}\) values much larger than those of the full-feature \textit{QDA}, whereas ordering by \(T_{j}\) resulted in substantial improvements relative to the full-feature \(\widehat{CER}\)s. The \textit{SIR-II} method exhibited the largest gain when it was reordered by the \(T_{j}\) criterion. In fact, the \(SIR\text{-}II_{T}\) method often achieved the minimum error rates. In contrast, the standard \textit{SIR-II} consistently produced \(\widehat{CER}\)s around 0.50. When \(n_{h} = p + 1 = 51\) for configuration Q1, the median \(\widehat{CER}\) achieved by the full-feature \textit{QDA} was 0.4287. However, regardless of sample size, the \(SIR\text{-}II_{T}\) method yielded an impressive median \(\widehat{CER}\) of approximately 0.02. In larger-sample scenarios, the \(T_{j}\)-ordered \textit{SDR} methods performed as well as or, in some cases, significantly better than the full-feature \textit{QDA}. For instance, when \(n_{h} = 5p = 250\) for configuration Q2, the \(SAVE_{T}\), \(SIR\text{-}II_{T}\), and \(SSDR_{T}\) methods achieved the lowest error rates, with \(SSDR_{T}\) performing best. For configuration Q3 with \(n_{h} = 250\), the \(\textit{PCA}_{T}\) method achieved the minimum median \(\widehat{CER}\) of 0.3484, compared to 0.4472 for \textit{QDA} and 0.4994 for standard \textit{PCA}. In contrast, all other \textit{SDR} methods yielded equal or higher error rates than the full-dimensional \textit{QDA}.

Under \textit{LDA} from Figure \ref{fig:3_LDA}, the \(T_{j}\)-ordered \textit{SDR} methods clearly achieved superior performance compared to their eigenvalue-ordered counterparts. In configuration L1, \textit{PCA} and \(PCA_{T}\) performed similarly. However, in configuration L2, \textit{PCA} produced a consistent median \(\widehat{CER}\) of approximately 0.50. For \(n = p + 1 = 51\) and \(n = 2p = 100\), \(PCA_{T}\) yielded the minimum error rates with median \(\widehat{CER}\)s around 0.0950 compared to the median \(\widehat{CER}\)s for \textit{LDA} of 0.1820 and 0.1235, respectively. Under Model \ref{eq:model}, when the parameters are known, \textit{LDA} yields the Bayes optimal subspace. Thus, for large samples, the full-feature \textit{LDA} achieved equal or better performance than \textit{LDA} did after we used the \textit{SDR} methods to reduce the feature space, which was expected. However, in the large-sample scenarios, ordering subspaces by \(T_{j}\) resulted in performance nearly identical to the optimal full-feature \textit{LDA} whereas eigenvalue ordering produced significantly larger \(\widehat{CER}\)s for certain methods across the configurations.

\subsection{Simulation for Continuous Response}
Through \textit{MC} simulations, we illustrated the efficacy of reordering the \textit{DRS} by using the \(F_j\) criterion when the response was continuous. We used the same \textit{SDR} methods as in Section \ref{subsec:5.1_CER_sims} and, likewise, referred to the \(F_j\)-ordered \textit{SDR} methods as \(PCA_{F}\), \(SAVE_{F}\), \(SIR\text{-}II_{F}\), and \(SSDR_{F}\). The eigenvalue-ordered \textit{SDR} methods are referred to by their original names. For all \textit{SDR} methods, we set the number of slices to \(H = 5\).

\begin{figure}[t!]
    \centering
    \includegraphics{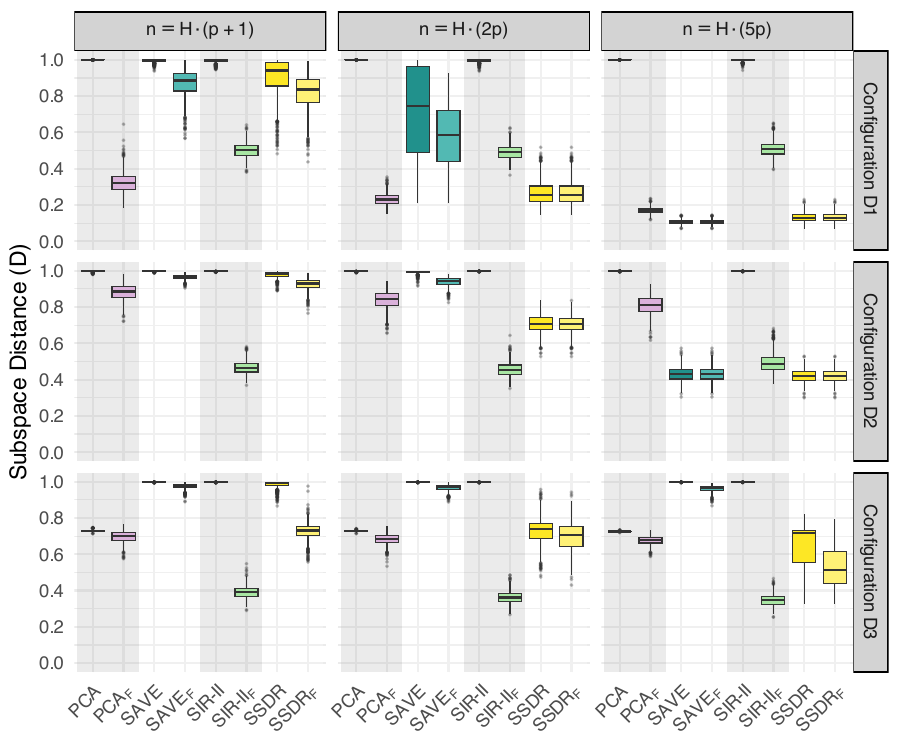}
    \caption{Estimated subspace distance \(\mathcal{D}\), given in \eqref{eq:D}, plots for contrasting eigenvalue-ordered \textit{SDR} methods (labeled by their standard names) with their \(F_{j}\)-ordered counterparts (denoted by the superscript \(F\)). The number of slices was set to \(H = 5\) for all \textit{SDR} methods.}
    \label{fig:4_D}
\end{figure}

Similar to our simulation set up in Section \ref{subsec:5.1_CER_sims}, we set \(p = 50\) for all parameter configurations. We varied the three training sample sizes of  \(n = H\cdot(p + 1)\), \(H\cdot(2p)\), and \(H\cdot(5p)\) to again simulate poorly-posed to well-conditioned estimation of the conditional moments within each slice. For each \textit{SDR} method, to evaluate subspace estimation accuracy and, hence, predictor–response preservation, we recorded the estimated subspace distance \(\mathcal{D}\) in \eqref{eq:D} between the true basis \(\boldsymbol{\beta} \in \mathbb{R}^{p \times d}\) and the estimated basis \(\widehat{\boldsymbol{\beta}} \in \mathbb{R}^{p \times d}\). We replicated this process 1,000 times for each parameter configuration and summarized the results in Figure \ref{fig:4_D}, which displays the distribution of \(\mathcal{D}\) for each \textit{SDR} method.

The configurations used are similar to those in \cite{lin2019} and \cite{zeng2024}. For configurations D1 and D2 below, \(\mathbf{X}_{i}\) follows a multivariate normal distribution with \(\boldsymbol{\mu} = \mathbf{0}_{p}\) and \(\boldsymbol{\Sigma} = AR(0.50)\), where \(AR(0.50)\) is a \(p \times p\) auto-regressive matrix whose \((i, j)^{\text{th}}\) element is \(0.50^{|i - j|}\). For each model, \(\varepsilon_{i}\) follows the \(\mathcal{N}(0, 1)\) distribution. The configurations are as follows. 
\begin{itemize}
    \item Configuration D1: \(Y_{i} = \boldsymbol{\beta}^{\top}\mathbf{X}_{i} + \varepsilon_{i}\), where 
    % \(\boldsymbol{\Sigma} = AR(0.5)\) and 
    \(\boldsymbol{\beta} \in \mathbb{R}^{p}\) with \textit{IID} entries from \(\mathcal{N}(0, 1)\). 
    \item Configuration D2: \(Y_{i} = (\boldsymbol{\beta}_{1}^{\top}\mathbf{X}_{i})\cdot\exp\left(\boldsymbol{\beta}_{2}^{\top}\mathbf{X}_{i} + \varepsilon_{i}\right)\), where 
    % \(\boldsymbol{\Sigma} = AR(0.5)\), 
    \(\beta_{1i}\)'s and \(\beta_{1j}\)'s have \textit{IID} entries from the uniform(0.30, 0.60) distribution for \(1\leq i \leq 30\) and \(1 \leq j \leq 15\), \(\beta_{1j}\)'s  have \textit{IID} entries from the uniform(-0.30, -0.60) for \(16 \leq j \leq 30\), and \(\beta_{ij} = 0\) otherwise. 
    \item Configuration D3: The same model as configuration D2, except \(\mathbf{X}_{i}\) has a non-elliptical distribution such that \(\mathbf{X}_{i} \sim 0.40\mathcal{N}(\boldsymbol{\mu}_{1}, \boldsymbol{\Sigma}_{1}) + 0.20\mathcal{N}(\boldsymbol{\mu}_{2}, \boldsymbol{\Sigma}_{2}) + 0.40\mathcal{N}(\boldsymbol{\mu}_{3}, \boldsymbol{\Sigma}_{3})\), where \(\boldsymbol{\mu}_{1} = (-\mathbf{1}_{30}, \mathbf{0}_{p - 30})\), \(\boldsymbol{\Sigma}_{1} = AR(0.10)\), \(\boldsymbol{\mu}_{2} = \mathbf{0}_{p}\), \(\boldsymbol{\Sigma}_{2} = AR(0.50)\), \(\boldsymbol{\mu}_{3} = -\boldsymbol{\mu}_{1}\), and \(\boldsymbol{\Sigma}_{3} = AR(0.90)\). 
\end{itemize}

From Figure \ref{fig:4_D}, we found that the \(F_{j}\)-ordered \textit{SDR} methods achieved superior subspace estimation compared to the eigenvalue-ordered counterparts. For every \textit{SDR} method, configuration, and sample size, the \(F_{j}\)-ordering resulted in either comparable performance to eigenvalue ordering or, in most cases, significantly lower \(\mathcal{D}\) values. For configuration D1, which was a single-index model with \(d = 1\), \textit{PCA} resulted in a median \(\mathcal{D}\) of 1.00, regardless of sample size, which indicated that no predictor–response information was preserved. This result is not surprising because \textit{PCA} considers only \(\mathbf{X}\) and is an unsupervised \textit{SDR} method. In contrast, for \(n = H\cdot(p + 1) = 55\), \(H\cdot(2p) = 500\), and \(H\cdot(5p) = 1{,}250\), \(PCA_{F}\) achieved median \(\mathcal{D}\) values of 0.3206, 0.2303, and 0.1692, respectively. Moreover, for \(n = 55\) and 500, \(PCA_{F}\) yielded the minimum \(\mathcal{D}\) values when contrasted to all other methods. This difference illustrated that \textit{PCA}, traditionally an unsupervised method, can become a pseudo-supervised method when it incorporates the \(F_{j}\)-ordering criterion, thus enabling it to compete with or outperform traditional supervised \textit{SDR} methods.

For all configurations, the \textit{SAVE} and \textit{SDRS} methods performed poorly for small sample sizes whereas the \(SAVE_{F}\) and \(SDRS_{F}\) methods yielded modest to substantial improvements. With larger sample sizes, the performances of \textit{SAVE} and \textit{SDRS} were similar across orderings with modest advantages for the \(F_{j}\)-based variants. Configurations D2 and D3 were multiple-index models with \(d = 2\). Here, similar to \textit{PCA}, the \textit{SIR-II} method yielded no subspace recovery when ordered by eigenvalue magnitude whereas the \(SIR\text{-}II_{F}\) method achieved superior predictor–response preservation, often yielding the minimum \(\mathcal{D}\) values.

% For sample sizes that are small relative to the feature-space dimension \(p\), we find that the empirical \(T_j\) or \(F_{j}\) values may prioritize directions that deviate from the population-optimal ordering. This occurs particularly when eigenvectors are poorly estimated due to relatively small sample sizes. However, from the simulation studies presented in in Section \ref{sec:5_simulation}, we found these directions significantly reduced error rates compared to ordering the subspaces by the eigenvalue magnitudes. This reflects the fact that \(T_j\) and \(F_{j}\) serve as a data-driven subspace criteria that corrects, to some extent, for subspace estimation variability and emphasizes maximum population/slice separation. However, as the sample size increases, the estimated subspace stabilizes, and the \(T_j\) and \(F_{j}\)-based ranking converge to the population-optimal ordering as shown in Theorem \ref{thm:T_rank} and \ref{thm:F_rank}, respectively. 

\section{Real Data Applications} \label{sec:6_real_data_apps}
In this section, we present an application of our proposed \(T_{j}\) subspace criterion to high-dimensional gene expression data and an application of our \(F_{j}\) criterion to continuous response data for housing price prediction. For simplicity of presentation, we considered two-fold validation for both data applications. Additional real-data applications are provided in the Supplementary Material, including several repeated 10-fold cross-validation error rate comparisons for each \textit{SDR} method using both \(T_{j}\) and \(F_{j}\), similar to those in Section \ref{subsec:5.1_CER_sims}. Moreover, we provide results for the \textit{DR} method within our proposed subspace ordering framework. We also include an analysis of brain cancer data illustrating the relationship between the empirical distribution of \(T_j\) and the empirical error rate distribution.

\subsection{Application to High-Dimensional Gene Expression Data} \label{subsec:6.1_golub}

Here, we present our analysis of high-dimensional gene expression data for two leukemia subtypes from \cite{golub1999}. The data were obtained from bone marrow samples of 72 patients: 47 diagnosed with \textit{acute lymphoblastic leukemia} (\textit{ALL}) and 25 with \textit{acute myeloid leukemia} (\textit{AML}). Affymetrix Hgu6800 chips were used to extract 7,129 gene expression levels from each patient.

\begin{figure}[t!]
    \centering
    \includegraphics{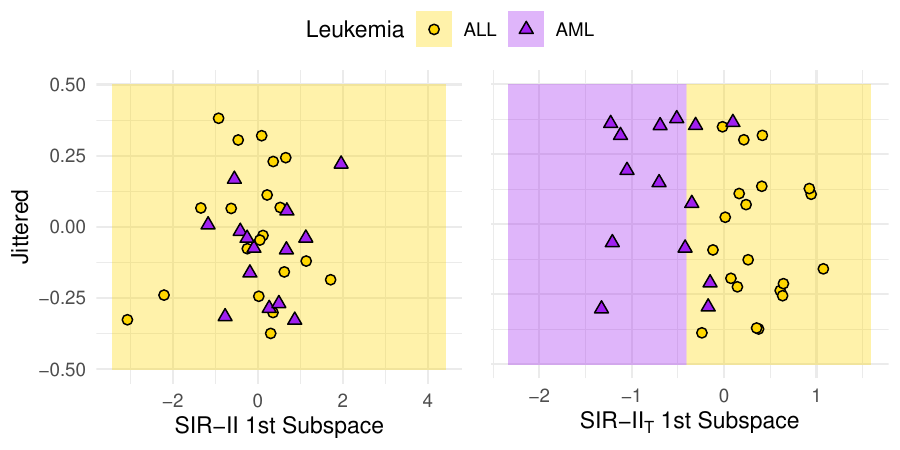}
    \caption{Leukemia data discussed in Section \ref{subsec:6.1_golub} in the estimated  \textit{SIR-II} subspaces corresponding to the largest eigenvalue (left plot) and the largest \(T_{j}\) value (right plot). \textit{QDA} was used for the estimated decision boundaries. Vertical jitter was added for visualization.}
    \label{fig:5_golub}
\end{figure}

The authors provided training and testing sets consisting of 38 and 34 patients, respectively. For the training set, we estimated the \textit{DRS} using the \textit{SIR-II} method. To ensure the generalized eigenvalue problem was solvable, we applied Tikhonov regularization to the sample covariance matrix such that  \(\widehat{\mathbf{N}} = \mathbf{S} + \gamma\mathbf{I}_{p}\) with \(\gamma \coloneqq 10^{-6}\). We obtained the \(SIR\text{-}II_{T}\) \textit{DRS} by reordering the \textit{SIR-II} basis using the \(T_{j}\) criterion. The dimensionality of the training and testing data was reduced to \(d = 1\), which we determined via validation using the training set. Next, we estimated the \textit{QDA} decision boundary using the reduced training data in both \textit{DRS}s. Our results are shown in Figure \ref{fig:5_golub}, which displays the reduced testing data in the estimated \textit{SIR-II} and \(SIR\text{-}II_{T}\) subspaces, respectively, along with the corresponding \textit{QDA} decision boundaries derived from the reduced training data.

From Figure \ref{fig:5_golub}, we readily see that the \textit{SIR-II} subspace associated with the largest eigenvalue yielded no discriminatory information. That is, no response information was preserved, and all patients were classified as \textit{ALL} by \textit{QDA}. This result yielded a \(\widehat{CER} = 0.4117\). In contrast, the \textit{SIR-II} subspace associated with the largest \(T_{j}\) achieved clear separation between the \textit{ALL} and \textit{AML} samples, with a substantially lower \(\widehat{CER} = 0.1471\). Every patient with \textit{ALL} was correctly identified, and only five patients with \textit{AML} were misclassified. Similar results were obtained under \textit{LDA}, where \textit{SIR-II} performed comparably, and \(SIR\text{-}II_{T}\) yielded a slightly higher \(\widehat{CER} = 0.1764\).

We attributed the superior performance of the \(SIR\text{-}II_{T}\) method to the \(T_{j}\) criterion, which provided a data-driven measure that 
% incorporated predictor–response information and 
yielded a more stable ordering criterion than eigenvalue magnitude did. In particular, due to singularity issues, the largest eigenvalue was \(10^{4}\) and non-unique across the first 7,092 eigenvectors. In contrast, the \(T_{j}\) values exhibited clear separation, where the largest value was 5.41, and the second largest was 1.44. The remaining \(T_{j}\) values decreased gradually and were unique across all subsequent subspaces. Figures showing the eigenvalue and \(T_{j}\) orderings are provided in the Supplementary Material. The eigenvector associated with the largest eigenvalue yielded \(T_{1} = 0.01\), ranking 6,968th among all \(T_{j}\) values whereas the subspace that achieved the largest \(T_{j} = 5.41\) corresponded to the smallest eigenvalue of 0.0001. Thus, under the traditional eigenvalue-ordering framework, the most informative subspace would not have been selected; however, using the \(T_{j}\) criterion, we identified this subspace as the most informative and stable direction.

\subsection{Application to Real Data with a Continuous Response}\label{subsec:6.2_boston}

In this section, we analyzed the Boston housing price data, originally compiled from the 1970 U.S. Census. The dataset contained 506 observations, each representing a census tract within the Boston metropolitan area. For each tract, 13 various economic and environmental characteristics were recorded, such as the per capita crime rate by town, average number of rooms per dwelling, full-value property tax rate per \$10,000, and others. The response variable was the median value of owner-occupied homes per \$1,000. See \cite{harrison1978} for additional details. Consistent with previous studies (e.g.,  \cite{chen2010}), we removed the observations with crime rates larger than 3.2 because they coincided with constant predictors. Thus, the data used for analysis had 374 observations. 

\begin{figure}[t!]
    \centering
    \includegraphics{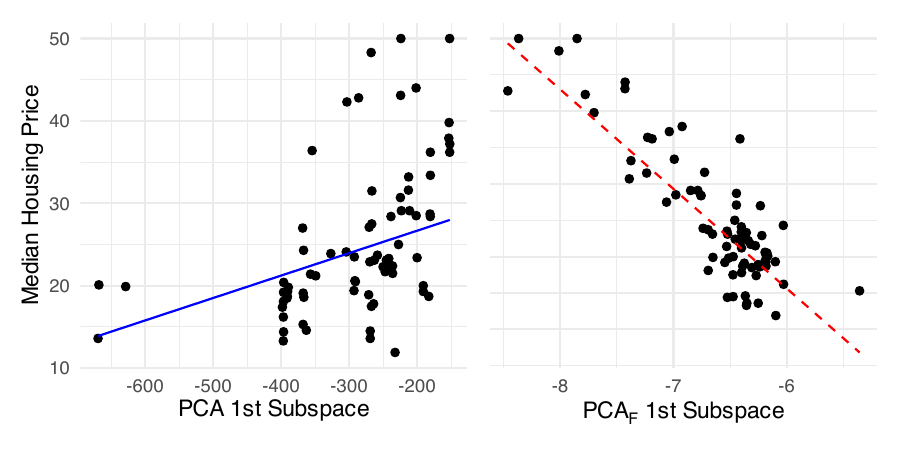}
    \caption{Boston housing data discussed in Section \ref{subsec:6.2_boston} in the estimated \textit{PCA} subspaces corresponding to the largest eigenvalue (left plot) and the largest \(F_{j}\) value (right plot). Estimated simple ordinary least squares solution using the training data is given by the line.}
    \label{fig:6_boston}
\end{figure}

We randomly partitioned the data into training (80\%) and testing (20\%) subsets for model estimation and evaluation. As in Section \ref{subsec:6.1_golub}, we estimated the \textit{DRS} using the training data for both the \textit{PCA} and \(PCA_{F}\) methods, thereby reducing the data dimensionality to \(d = 1\), which we determined via the training data. Figure \ref{fig:6_boston} presents our results, displaying scatter plots between the median housing prices for the testing set and the reduced testing data in the estimated \textit{PCA} and \(PCA_{F}\) subspaces, respectively. From Figure \ref{fig:6_boston}, we again found that the subspace associated with the largest eigenvalue captured very little response information. In the traditional first \textit{PCA} subspace, the resulting testing \textit{mean squared error} (\textit{MSE}) from a simple linear regression model was 67.64. 
% The model was clearly insignificant and yielded an \textit{MSE} comparable to that obtained when we used only the training sample mean, which we found to be 62.82. 
In contrast, the \(F_{j}\) criterion identified a subspace, specifically the 11th \textit{PCA} direction, that captured a meaningful predictor-response structure and reduced the testing \textit{MSE} to 22.12. 

This real-data application highlights an important limitation of ordering by eigenvalue magnitude. Although larger eigenvalues typically correspond to directions with greater variability in the data and are often assumed to capture more information, this assumption generally does not hold in practice. In this application, the data did in fact exhibit greater overall variability in the \textit{PCA} subspace associated with the largest eigenvalue than in the first \(PCA_{F}\) subspace. However, this increased variability arose because the eigenvector corresponding to the largest eigenvalue was sparse and effectively projected the data onto two distinct spans. In contrast, the \(F_{j}\) criterion considered meaningful variability relative to the response and predictors, which yielded a subspace that captured a more informative structure than just incidental variance.

\section{Discussion}\label{sec:discussion}

We proposed using criteria other than eigenvalue magnitude for ordering subspaces in \textit{SDR}. Traditionally, researchers have used eigenvalues to determine the basis for a \textit{DRS}; however, we have demonstrated that eigenvalues generally do not correspond to the predictive relevance of a subspace. Therefore, we proposed and established theoretical results for the \(T_j\) and \(F_j\) criteria, which rank subspaces by their relevance to the response. The proposed criteria provide a straightforward and interpretable alternative that aligns the subspace ordering step with the goal of \textit{SDR}---namely, maximizing predictor-response preservation.

Although we do not claim that reordering subspaces by our proposed criteria universally yields the best predictive performance, we have found considerable evidence through both simulations and real-data applications that ordering a \textit{DRS} by the \(T_j\) and \(F_j\) criteria generally results in lower misclassification rates and more accurate subspace estimation than ordering by eigenvalue magnitude. Specifically, in high-dimensional or ill-conditioned settings, eigenvalues can yield an unstable subspace criterion with non-unique values. Moreover, when the leading eigenvectors are sparse, the associated eigenvalues no longer reflect response-related variability but instead capture incidental variance. In contrast, our proposed criteria provide a stable and superior ordering of subspaces. This fact reflects that \(T_j\) and \(F_j\) serve as data-driven subspace criteria that, to some extent, correct for subspace estimation variability by emphasizing maximum population or slice separation. Additionally, ordering subspaces by either \(T_j\) or \(F_j\) enables unsupervised methods, such as \textit{PCA}, to behave in a pseudo-supervised manner and, in some cases, compete with or even outperform conventional supervised \textit{SDR} methods.

We view this work as an initial step toward developing a broader class of prediction-specific subspace ordering criteria. Our goal is to have demonstrated through the proposed \(T_{j}\) and \(F_{j}\) criteria that subspace ordering can and often should be informed by the objectives of the supervised learner rather than defaulting to eigenvalue magnitude. In this paper, we considered discrimination and regression settings; however, deriving alternative subspace criteria tailored to other learning objectives remains for future research. Moreover, we assumed that the structural dimension of the \textit{CS} was known; jointly estimating the structural dimension while simultaneously ordering subspaces using a predictive importance measure is another related future research topic. Such extensions would allow \textit{SDR} methods to integrate more tightly with the intended supervised model and produce a \textit{DRS} that is not only statistically sufficient but also optimally informative for prediction. 

\subsection*{Code Availability}
All reproducible code for the simulations, real data applications, and graphical figures is available at
% [github link removed for anomyziation]
\url{https://github.com/DerikTBoonstra/Subspace_Ordering}
. Moreover, the \(T_{j} \) and \(F_{j}\) criteria, along with all of the \textit{SDR} methods used, are implemented in the working 
% [redacted]
\texttt{sdr} 
\verb|R| package available at 
% [github link removed for anomyziation]
\url{https://github.com/DerikTBoonstra/sdr}
.

\subsection*{Acknowledgements} 
\noindent The authors received no financial support from any funding agency in the public, commercial, or not-for-profit sectors.

\hypersetup{bookmarksdepth=-1}
%\bibliography{bibliography.bib}

\end{document}